\documentclass[aps,prb,twocolumn,superscriptaddress,floatfix]{revtex4-1}
 
\usepackage{latexsym,amssymb,float}
\usepackage{graphicx}
\usepackage{amsmath}
\usepackage{bm}
\usepackage{tikz}
\usepackage{booktabs}
\usepackage[breaklinks=true,unicode=true,urlcolor = blue,colorlinks = true,citecolor = blue,linkcolor = blue]{hyperref}

\newcommand{\be}{\begin{equation}}
\newcommand{\ee}{\end{equation}}
\renewcommand{\vec}[1]{\bm{#1}}

\renewcommand{\vr}{\vartheta} % velocity ratios
\newcommand{\tu}{\tilde{u}} % transformed u
\newcommand{\tl}{\tilde{\lambda}} % transformed \lambda

\begin{document}

\title{Quantum criticality on a compressible lattice}
\author{Saheli Sarkar$^*$}
\affiliation{Institute for Quantum Materials and Technology, Karlsruhe Institute of Technology, D-76131 Karlsruhe, Germany}

\author{Lars Franke$^*$}
\affiliation{Institute of Theoretical Solid State Physics, Karlsruhe Institute of Technology, D-76131 Germany}

\author{Nikolas Grivas}
\affiliation{Institute of Theoretical Solid State Physics, Karlsruhe Institute of Technology, D-76131 Germany}

\author{Markus Garst}
\affiliation{Institute for Quantum Materials and Technology, Karlsruhe Institute of Technology, D-76131 Karlsruhe, Germany}
\affiliation{Institute of Theoretical Solid State Physics, Karlsruhe Institute of Technology, D-76131 Germany}

\begin{abstract}
The stability of a quantum critical point in the $O(N)$ universality class with respect to an elastic coupling, that preserves $O(N)$ symmetry, is investigated for isotropic elasticity in the framework of the renormalization group (RG) close to the upper critical dimension $d=3-\epsilon$. With respect to the Wilson-Fisher fixed point, we find that the elastic coupling 
is relevant  in the RG sense for $1\leq N \leq 4$, and the crystal becomes microscopically unstable, i.e., a sound velocity vanishes at a finite value of the correlation length $\xi$. For $N > 4$,  an additional fixed point emerges that is located at a finite value of the dimensionless elastic coupling. This fixed point is repulsive and separates the flow to weak and strong elastic coupling. As the fixed point is approached the sound velocity is found to vanish only asymptotically as $\xi \to \infty$ such that the crystal remains microscopically stable for any finite value of $\xi$.
The fixed point structure we find for the quantum problem is distinct from the classical counterpart in $d=4-\epsilon$, where the crystal always remains  microscopically stable for finite $\xi$.
\\[0.5em]
*These authors contributed equally.
\end{abstract}

\pacs{}

\maketitle

\section{Introduction}

In the solid state, any degree of freedom invariably interacts with the low-energy fluctuations of the atomic crystal lattice, i.e., the acoustic phonons. Often, this interaction does not lead to qualitatively new behaviour as it is effectively weak. Notable exceptions are materials close to certain second-order phase transitions. Here, it is important to distinguish whether the symmetry allows a bilinear coupling between strain and the order parameter of the 
phase transition or not. In the former case, the order parameter hybridises with the elastic degrees of freedom and the critical behaviour is strongly affected by critical elasticity \cite{Cowley1976,Zacharias2015,Zacharias2015-2}. Examples include piezoelectric ferroelectricity \cite{Villain1970,Levanyuk1970}, Mott criticality \cite{Zacharias2012,Gati2016}, metamagnetic criticality \cite{Weickert2010} or nematic quantum criticality \cite{Paul2017,Reiss2020}.

In case that the bilinear coupling is not allowed the situation is more intricate. For classical criticality, this problem was intensively investigated from the 1950ies to the 70ies \cite{Rice1954,Domb1956,Mattis1963,Fisher1968,LarkinPikin1969,Rudnick1974,Sak1974,Wegner1974,BergmanHalperin1976,DeMoura1976,Nattermann1977,BrunoSak1980}, see Ref.~\onlinecite{Duenweg2000} for a review. 
It was pointed out by Rice already in 1954  that a critical system with a diverging specific heat should exhibit a non-perturbative elastic coupling\cite{Rice1954}.
In the absence of any common symmetries, the strain tensor $\varepsilon_{ij}$ might only couple linearly to the square of the order parameter, i.e., the energy density of the critical degrees of freedom. Perturbatively integrating out the latter leads to a renormalization of the elastic constants proportional to their equal-time energy density autocorrelations, i.e, the critical specific heat. As a consequence, this perturbative treatment necessarily breaks down if the critical specific heat, $C_{\rm cr} \sim |T-T_c|^{-\alpha}$ diverges with a positive exponent $\alpha > 0$ as the temperature $T$ approaches the critical temperature $T_c$. Assuming hyperscaling $\alpha = 2 - \nu d$ with the correlation exponent $\nu$ and the spatial dimension $d$, the criterion for a non-perturbative elastic coupling amounts to $2 - \nu d > 0$.

In order to determine the fate of the system for $\alpha > 0$, it is essential to take into account both the elastic moduli as well as the local rigidity of the crystal, which must be all positive to ensure, respectively, its macroscopic and microscopic elastic stability\cite{Landau1986}. Importantly, the strain tensor $\varepsilon_{ij}$ comprises a uniform part,  
$E_{ij} \propto \int d\vec r\, \varepsilon_{ij}$, 
as well as a non-uniform part carrying finite wavevector $\vec q$. The eigenmodes of 
$E_{ij}$ are determined, on the one hand, by the elastic constant tensor $C_{ijkl}$ whose eigenvalues are the elastic moduli. The elastic rigidities, on the other hand, are given by the acoustic phonon velocities, which are obtained via  the eigenvalues of the dynamical matrix $\mathcal{D}_{ik} = q_j q_l C_{ijkl}$ for wavevectors $\vec q$. Elastic coupling to criticality poses the challenge that infinite-range interactions among the critical degrees of freedom are generated in both cases, when integrating out the macroscopic strain $E_{ij}$ as well as when integrating out the non-uniform phonon degrees of freedom.

This challenge was  successfully addressed first by Larkin and Pikin \cite{LarkinPikin1969} for an isotropic elastic system. Using a Hubbard-Stratonovich transformation, they succeeded to decouple the infinite-range interactions  yielding an effective non-analytic Landau theory for a macroscopic auxiliary field. The result for $\alpha > 0$ crucially depends on the imposed boundary conditions: the phase transition turns from second- to first-order at constant hydrostatic pressure $P$, whereas at constant volume $V$ a modified critical point might be reached that is characterized by Fisher-renormalized exponents\cite{Fisher1968}. In particular, the critical part of the specific heat behaves in this case $C_{\rm cr} \sim |T-T_c|^{-\alpha_F}$ with $\alpha_F = -\alpha/(1-\alpha) < 0$. 

Subsequent works essentially confirmed this scenario but further elucidated the subtle importance of boundary conditions\cite{Wegner1974,DeMoura1976,BergmanHalperin1976}. Bergman and Halperin\cite{BergmanHalperin1976} pointed out that the boundary condition of constant volume in an isotropic elastic system might not be sufficient to reach the Fisher-renormalized critical point. Instead, pinned boundary conditions need to be imposed, that fix each unit cell of the crystal at the surface, and, in addition, an internal fracture of the sample must be prevented. 
 
It is interesting how this physics is captured in the framework of a renormalization group (RG) analysis. In the literature, two distinct approaches can be found treating a classical critical $\phi^4$ theory with $O(N)$ symmetry in $d = 4-\epsilon$ dimensions and isotropic elasticity. Similar to the work of Larkin and Pikin\cite{LarkinPikin1969}, the first approach considers an effective theory for criticality after integrating out all elastic degrees freedom at the cost of infinite-range interactions\cite{Rudnick1974,Sak1974,DeMoura1976,BrunoSak1980}. For $\epsilon > 0$ and in the absence of an elastic coupling the RG trajectories are governed by the stable Wilson-Fisher (WF) fixed point with a specific heat exponent $\alpha = \epsilon (4-N)/(2 (N+8))$ whose sign depends on the number of order parameter components $N$. Consistent with expectations, the WF fixed point becomes unstable with respect to the elastic coupling for $N < 4$ when $\alpha > 0$. Depending on the applied boundary conditions, the RG flow is either towards a stable Fisher-renormalized Wilson-Fisher (FR-WF) fixed point for constant volume $V$ or one finds runaway flow for constant pressure $P$, that is interpreted as a sign for a first-order transition\footnote{In Refs.~\onlinecite{Sak1974,BrunoSak1980} only free boundary conditions at constant $P$ were considered and the FR-WF fixed point was inaccessible.}. 

The second approach treats the critical and elastic degrees of freedom on the same footing\cite{BergmanHalperin1976} leading to a more transparent interpretation. Here, one also finds for $\epsilon > 0$ both fixed points, WF and FR-WF, whereas the former becomes  unstable again for $N < 4$. In this approach, the microscopic degrees of freedom, that includes the elastic rigidity of the phonon modes, then always flow towards the stable FR-WF fixed point. Upon approaching this fixed point,  the velocity of longitudinal phonons asymptotically vanishes as a function of the correlation length $\xi$ with a power law. Microscopically, the crystal thus only destabilizes asymptotically for $\xi \to \infty$ but remains stable at any finite distance to the FR-WF fixed point. In contrast, the macroscopic stability is determined by the bulk modulus that is found to reach zero at a finite distance to the FR-WF fixed point. Whether the macroscopic instability develops or not depends however again on the imposed boundary conditions. Moreover, if elastic anisotropies are taken into account, the crystal becomes microscopically unstable at a finite value of $\xi$ leading to a first-order transition irrespective of the boundary conditions.

It is an obvious question in which sense these results generalize to quantum criticality. It is rather straightforward to obtain the criterion for a non-perturbative elastic coupling to quantum criticality\cite{Anfuso2008,Zacharias2015-2}. Generally, close to a second-order quantum critical point the ground state energy varies as a function of the tuning parameter $r$ as $\delta E_0 \sim |r|^{2-\alpha_q}$  with the exponent $\alpha_q$. If the tuning parameter either depends on pressure or volume, this amounts to a correction to either the compressibility or the bulk modulus proportional to $\partial^2_r \delta E_0$ that diverges for $r\to 0$ if $\alpha_q > 0$. The exponent $\alpha_q$ is thus the quantum analogue of the specific heat exponent $\alpha$ for classical criticality. Assuming hyperscaling with the dynamical critical exponent $z$ and the correlation length exponent $\nu$ of the quantum phase transition, the criterion for a non-perturbative elastic coupling to quantum criticality amounts to 
\begin{align} \label{Criterion}
\alpha_q = 2 - \nu (d+z) > 0.
\end{align}
This can be also formally obtained from the classical criterion by replacing $d \to d+z$. 

In Ref.~\onlinecite{Anfuso2008} it was pointed out that the criterion \eqref{Criterion} might be fulfilled at quantum critical points of low-dimensional spin systems. Moreover, the critical temperature for the elastically-induced, putative first-order transition was estimated for the field-driven quantum critical point of the spin-ladder compound (C$_5$H$_{12}$N)$_2$CuBr$_4$. The criterion is also marginally fulfilled at a quantum Lifshitz transition in $d=2$ and, indeed, a large lattice softening as a function of temperature was recently observed at the two-dimensional Van Hove singularity of Sr$_2$RuO$_4$ \cite{Noad2023}. An interesting scenario of a quantum annealed criticality was proposed in Ref.~\onlinecite{Chandra2020}, which involves a quantum critical point that does not fulfil the criterion \eqref{Criterion} and stays second-order but terminates a line classical phase transitions, which are converted to first-order by elastic coupling.

A first serious study going beyond perturbation theory and considering quantum criticality in the presence of a coupling to isotropic elasticity was carried out by Chandra, Coleman, Continentino and Lonzarich \cite{Chandra2020}. They considered a Lorentz invariant Ising $\phi^4$ theory with dynamical exponent $z=1$ that exhibits a Wilson-Fisher (WF) fixed point in $d = 3-\epsilon$ dimensions for $\epsilon > 0$. They followed the approach of Larkin and Pikin\cite{LarkinPikin1969} and first integrated out all elastic degrees of freedom. This results in an effective quantum theory for the order parameter that contains $(i)$ an infinite-range interaction in space-time and $(ii)$ an interaction that is non-analytic in frequency-wavevector space. 
From the latter $(ii)$, only the zero frequency limit was retained that just renormalizes the local interaction 
while the rest is disregarded claiming that it is irrelevant with respect to the WF fixed point.
Remarkably, the authors find that the interaction $(i)$ is not only of infinite-range in space as in the classical problem but simultaneously of infinite-range in time. Decoupling this infinite-range interaction via a Hubbard-Stratonovich transformation, the authors arrive at a non-analytic Landau theory for an auxiliary field generalizing the result of Larkin and Pikin to the quantum realm. This suggests, in particular for $\epsilon > 0$, the existence of a quantum version of a Fisher-renormalized Wilson-Fisher (FR-WF) fixed point.

Another investigation of the $O(N)$ symmetric version of the same model was recently performed by Samanta, Shimshoni and Podolsky \cite{Samanta2022}. Elastic degrees of freedom were also integrated out but the infinite-range interaction $(i)$, which is key to this problem, was now omitted. The interaction $(ii)$, that is non-analytic in frequency-wavevector space, was expanded in terms of a spherical harmonic decomposition and truncated at some order. The resulting theory was analysed in terms of a RG treatment in $d = 3-\epsilon$ dimensions. It was found that the interaction $(ii)$, in contrast to the claim of Ref.~\onlinecite{Chandra2020}, destabilizes the Wilson-Fisher fixed point for $N<4$. 
 
Given these two contradicting theories, the fate of a quantum critical point in the presence of an elastic coupling 
is currently unsettled. Various questions arise: How does the quantum system behave if the criterion \eqref{Criterion} is fulfilled? Does the crystal necessarily undergo a first-order isostructural transition or can it be stabilized by pinned boundary conditions? 
Does a quantum version of a FR-WF fixed point exist as suggested by the results of Chandra {\it et al.} \cite{Chandra2020}? 
Could the crystal become unstable even if the perturbative criterion \eqref{Criterion} is not fulfilled? 
In order to shed light on these questions, we also consider in the present work the Lorentz invariant quantum critical $\phi^4$ theory with $O(N)$ symmetry and a coupling to isotropic elasticity. Following the approach of Bergmann and Halperin\cite{BergmanHalperin1976}, we perform a RG analysis of the quantum problem at temperature $T=0$ treating both critical and elastic degrees of freedom on the same footing.

The following sections are organized as follows. In section \ref{sec:II} the field theory is defined, and its infrared singularities are discussed in subsection \ref{subsec:LogSing}. The one-loop RG equations are presented in subsection \ref{subsec:RGEq} and discussed in subsection \ref{subsec:RGFlow}. The macroscopic stability of the crystal is analyzed in subsection \ref{sec:CrystalStability}. In Section \ref{sec:Discussion}  the results are summarized and discussed.

\section{Quantum critical $\phi^4$ theory coupled to isotropic elasticity}
\label{sec:II}

We consider an Euclidian field theory, $\mathcal{S} = \int_0^\beta d\tau \int d\vec r \, \mathcal{L}$, for the $N$ component vector field $\vec \phi(\vec r,\tau)$ 
and the displacement field $\vec u(\vec r, \tau)$ governed by the Lagrangian density 
$\mathcal{L} = \mathcal{L}_{\phi} + \mathcal{L}_{\varepsilon} + \mathcal{L}_{\phi-\varepsilon} $,
\begin{align}
\mathcal{L}_{\phi} &= \frac{1}{2} \Big[(\partial_\tau \vec \phi)^2 + c^2 (\partial_i \vec \phi)^2 + r \vec \phi^2\Big] + \frac{u}{4!} (\vec \phi^2)^2,\\
\mathcal{L}_{\varepsilon} &= \frac12 \rho \left(\partial_\tau \vec u\right)^2 + \frac12 \varepsilon_{ij} C_{ijkl} \varepsilon_{kl}, \\
\mathcal{L}_{\phi-\varepsilon} &= \lambda\,  \vec \phi^2\, {\rm tr}\{\varepsilon_{ij}\}.
\end{align}
The critical part $\mathcal{L}_{\phi}$ is given by the Lorentz invariant $\phi^4$ theory with velocity $c$, tuning parameter $r$ and self-interaction $u$. The elastic part $\mathcal{L}_{\varepsilon}$ depends on the mass density $\rho$ and the elastic constant tensor $C_{ijkl}$. For a cubic crystal, it possesses only three independent components, i.e., $C_{11}$, $C_{12}$ and $C_{44}$ in Voigt notation. They are related to the bulk modulus $K$ and shear modulus $\mu$,
\begin{equation}
K = \frac{1}{3} C_{11} + \frac{2}{3} C_{12}, 
\qquad \mu = C_{44}.
\end{equation}
The anisotropy of the cubic lattice can be quantified by the anisotropy index $A = 2C_{44}/(C_{11}-C_{12})$, and, 
in the following, we focus on an isotropic crystal $A = 1$.
The strain tensor  in terms of the displacement field is given by $\varepsilon_{ij}(\vec r,\tau) = \frac{1}{2} (\partial_i u_j(\vec r,\tau) + \partial_j u_i(\vec r,\tau))$. As already alluded to in the introduction, the uniform part 
%$E_{ij}(\tau) = 
$\int d\vec r\, \varepsilon_{ij}(\vec r,\tau)$ plays a particular role; its effective potential determines the macroscopic stability of the crystal. Finally, $\lambda$ is the elastic coupling that maintains $O(N)$ symmetry of the order parameter field.

At zero temperature and spatial dimensions $d = 3$, both the self-interaction $u$ as well as the elastic coupling $\lambda$ are marginal with respect to the Gaussian (G) fixed point at $r = 0$. For $\lambda = 0$, the theory for the order parameter possesses Lorentz invariance with dynamical exponent $z = 1$, and it is at its upper critical dimension $d+z = 4$. As a consequence, for $\epsilon = 3 - d > 0$ the coupling $u$ becomes relevant, and the Gaussian fixed point is unstable towards 
the Wilson-Fisher (WF) fixed point. In the following, we will determine the influence of a finite elastic coupling $\lambda$ on the RG flow. We will concentrate on the non-condensed phase $r \geq 0$ at zero temperature.
 
 \begin{figure}
\includegraphics[width=\columnwidth]{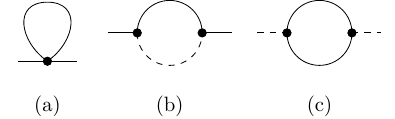}
\caption{Self-energy diagrams (a) and (b) for the order parameter field $\vec \phi$, and (c) for the phonon field $\vec u$. The solid and dashed lines are $\vec \phi$ and $\vec u$ propagators. 
 }
\label{fig:selfenergy_diagrams}
\end{figure}

The fluctuations of the order parameter around the Gaussian fixed point are characterized by the Green function
\begin{align}
G^{(0)}_{\alpha \beta}(\vec q, \omega) &= 
\frac{\delta_{\alpha \beta}}{\omega^2 + c^2 q^2 + r},
\end{align}
where $\alpha,\beta \in \{1,..,N\}$, $q = |\vec q|$ and $\omega$ is a bosonic Matsubara frequency. The fluctuations of the non-uniform strain component with $\vec q \neq 0$ are governed by the phonon Green function,
\begin{align} \label{PhononPropagator}
D^{(0)}_{ij}(\vec q, \omega) &= 
((\rho \omega^2 + \mathcal{D}(\vec q))^{-1})_{ij}
\nonumber\\
&= \frac{1}{\rho} \Big[\frac{\hat q_i \hat q_j}{\omega^2 + c_L^2 q^2} + \frac{\delta_{ij} - \hat q_i \hat q_j}{\omega^2 + c_T^2 q^2} \Big],
\end{align}
where $i,j \in \{1,2,3\}$ and $\hat q = \vec q/q$. The dynamical matrix $\mathcal{D}_{ik}(\vec q) = q_j q_l C_{ijkl}$, and the longitudinal and transversal sound velocities are given by $c_L = \sqrt{(K + 4\mu/3)/\rho}$ and $c_T = \sqrt{\mu/\rho}$, respectively. As the strain enters the elastic coupling only via its trace, only longitudinal phonons couple to the order parameter field. 

\subsection{Logarithmic singularities at one-loop order}
\label{subsec:LogSing}

In the following, we determine the logarithmic singularities generated by one-loop self-energy and vertex corrections in $d=3$. 
The self-energy diagrams are displayed in Fig.~\ref{fig:selfenergy_diagrams} where the solid line and dashed line  corresponds to a $G$ and $D$ Green function, respectively. The diagrams (a) and (b) contribute to the self-energy for the $\vec \phi$ field, $G^{-1}(\vec q,\omega) = (G^{(0)}(\vec q,\omega))^{-1} - \Sigma(\vec q,\omega)$, and diagram (c) contributes to the polarization for the phonons $D^{-1}(\vec q,\omega) = (D^{(0)}(\vec q,\omega))^{-1} - \Pi(\vec q,\omega)$. We obtain
\begin{align}
\label{eq:self-energy_phi}
\Sigma(\vec q, \omega) &= - 2 (2N + 4) \frac{u}{4!} I_{1} + 4 \lambda^2 I_2(\vec{q}, \omega),\\
\label{eq:self-energy_u}
 \Pi_{ij}(\vec{q},\omega) &= 2N \lambda^2 I_{3} q_i q_j ,
\end{align}
where the integrals $I_j$ with $j=1,2,3$ are specified below. We neglected already the wavevector and frequency dependence of $I_3(\vec q,\omega) \approx I_3(0,0) = I_3$ because it does not give rise to singular corrections.

\begin{figure}
    \centering
    \includegraphics[width=0.7\columnwidth]{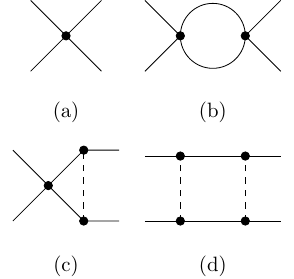}
        \caption{Vertex diagrams for the $\phi^4$ interaction. (a) is the bare vertex $u$ and (b)-(d) are vertex corrections. 
}
    \label{fig:phi4_vertex_diagrams}
\end{figure}

\begin{figure}
    \centering
        \includegraphics[width=0.9\columnwidth]{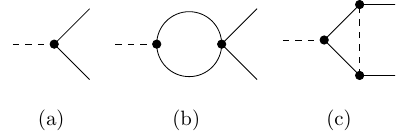}
    \caption{Vertex diagrams for the elastic coupling. (a) is the bare elastic coupling $\lambda$ and (b)-(c) are vertex corrections.    }
    \label{fig:elint_vertex_diagrams}
\end{figure}

The vertex diagrams for the self-interaction of the $\vec \phi$ field and the elastic coupling are listed in Figs.~\ref{fig:phi4_vertex_diagrams} and \ref{fig:elint_vertex_diagrams}, respectively. By power counting, singular corrections are only expected for the local interaction $u \to u + \delta u$ and $\lambda \to \lambda + \delta \lambda$ and we find
\begin{align}
\delta u &= - (4N + 32) \frac{u^2}{4!} I_3 + 24\lambda^2 u I_{4} - 96 \lambda^{4} I_5,\\
\delta\lambda &= - (4N + 8) \frac{u \lambda}{4!} I_3 + 4\lambda^3 I_4 ,
\end{align}
with additional integrals $I_j$ with $j = 4,5$.

The integrals are explicitly given by 
\begin{align} \label{Integral1}
 I_1 &= \int \frac{d\vec q d\omega}{(2\pi)^{d+1}}  
 \frac{1}{\omega^2 + c^2q^2 + r}, \\\label{Integral2}
 I_2(\vec q', \omega') &= \frac{1}{\rho}\int \frac{d\vec q d\omega}{(2\pi)^{d+1}}  
 \frac{1}{\omega^2 + c^2 q^2 + r} \\\nonumber &\qquad\quad \times
       \frac{(\vec{q'} +\vec{q})^2}{(\omega' + \omega)^2 + c_{L}^2(\vec{q'} + \vec{q})^2}.
\end{align}
Moreover, $I_3 = - \frac{d I_1}{d r}$, $I_4 = -\frac{d I_2(0,0)}{d r}$ and $I_5 = - \frac{1}{\rho }\frac{d I_4}{d c_L^2}$. We found it convenient to evaluate the integrals over the full frequency range but to introduce a UV cutoff $\Lambda$ for the wavevector integrals. Extracting the cutoff dependence from the integrals, we obtain
\begin{align}
I_1 &\simeq \frac{1}{8\pi^2 c^3} \left[c^2 \Lambda^2 - r \log \Lambda \right],\\
I_2 &\simeq \frac{1}{8 \pi^2 \rho} \frac{1}{c^3 c_L (c + c_L)}
\left[c^2 \Lambda^2 - \frac{2c + c_L}{c + c_L} r \log \Lambda \right] \nonumber \\
        &\quad- \frac{1}{4 \pi^2 \rho} \frac{1}{c c_L(c+ c_L)^3} \left( \omega^2 - \frac{1}{3} c^2 q^2 \right)  \log \Lambda.
\end{align}

Using these results, we will derive the renormalization group equations in the next section using a  Wilsonian scheme. The dependence of the integrals $I_i$ on $\Lambda^2$ will only influence the initial RG flow. In the following, we concentrate on the $\log \Lambda$ dependences that will control the flow at large wavelengths and low energies. We verified that the same one-loop RG equations are obtained by using instead a dimensional regularization scheme.

\subsection{Renormalization group equations}
\label{subsec:RGEq}

We perform a Wilsonian RG analysis by subsequently integrating out simultaneously $\vec \phi$ and $\vec u$ modes 
with large wavevectors. After integrating out a momentum shell $q \in [\Lambda/b,\Lambda]$ with $b > 1$ and $\log b \ll 1$ we rescale wavevectors and frequencies according to
\begin{align}
\vec q & = \vec q'/b, \quad \omega = \omega'/ b^z,
\end{align}
where we find it convenient to rescale frequencies with an arbitrary dynamical exponent $z$. In addition, the fields are rescaled, 
\begin{align}
\vec \phi(\vec q,\omega) &= \sqrt{Z_\phi} b^{\frac{d + 3 z}{2}} \vec \phi'(\vec q',\omega'),\\
\vec u(\vec q,\omega) &= \sqrt{Z_u} b^{\frac{d + 3 z}{2}} \vec u'(\vec q',\omega'),
\end{align}
where $Z_\phi$ and $Z_u$ are wavefunction renormalizations and $d$ is the spatial dimensionality. Finally, the parameters are rescaled according to
\begin{align}
c^2 & = c'^2/(Z_\phi b^{(2 z -2)}),\quad
c_L^2 = c'^2_L/(Z_u b^{(2 z -2)}),\nonumber\\
u &= u'/(Z_\phi^2 b^{(3 z-d)}),\quad
\lambda = \lambda'/(Z_\phi \sqrt{Z_u} b^{(5 z-d-2)/2}),\nonumber \\
r&= r'/(Z_\phi b^{2 z}).
\end{align}
Imposing the renormalization group conditions
\begin{align}
G^{-1}(0,\omega)|_{r = 0} = \omega^2, \quad
D^{-1}(\vec q,\omega)|_{r = 0} \overset{\vec q \to 0}{\to} \rho \omega^2,
\end{align}
we obtain the following set of differential RG equations up to one-loop order
\begin{widetext}
\begin{align}
    \label{eq:RG_eqn_Z}
 \frac{d \log Z_\phi}{d \ell} & =- \frac{1}{\pi^2 \rho} \frac{\lambda^2}{c c_L (c + c_L)^3},
 \\
  \label{eq:RG_eqn_r}
    \frac{d r}{d\ell}
        &= \bigg(2z - \frac{1}{\pi^2\rho} \frac{\lambda^2}{cc_L(c+c_L)^3}
        + \frac{1}{2\pi^2\rho} \frac{2c + c_L}{c_L(c+c_L)^2} \frac{\lambda^2}{c^3}
        - \frac{N+2}{48\pi^2} \frac{u}{c^3}\bigg) r  ,\\
\label{eq:RG_eqn_c}
    \frac{d c^2}{d \ell} 
        &= \left(2z - 2 - \frac{4}{3\pi^2\rho} \frac{\lambda^2}{cc_L(c+c_L)^3} \right) c^2 ,\\
    \label{eq:RG_eqn_cs}
    \frac{d c_L^2}{d \ell}
        &= \left(2z - 2 - \frac{N}{4\pi^2\rho} \frac{\lambda^2}{c_L^2c^3}\right)c_L^2 ,\\
      \label{eq:RG_eqn_u}
    \frac{d u}{d \ell}
        &= (3z - d)u
          - \frac{(N+8)}{48\pi^2} \frac{u^2}{c^3}
       + \frac{1}{\pi^2\rho} \frac{4 c^2 + 9 c c_L + 3 c_L^2}{c^3 c_L(c+c_L)^3} u \lambda^2
        - \frac{12}{\pi^2\rho^2} \frac{c^2 + 3cc_L + c_L^2}{c^3c_L^3(c+c_L)^3} \lambda^4 ,\\
    \label{eq:RG_eqn_l}
    \frac{d\lambda}{d\ell}
        &= \left(\frac{5z - d - 2}{2}
+       \frac{1}{2\pi^2\rho} \frac{3 c + c_L}{c^3 (c+c_L)^3} \lambda^2
        - \frac{N+2}{48\pi^2} \frac{u}{c^3} \right) \lambda ,
\end{align}
where we abbreviated $\ell = \log b$. In addition, we find $d \log Z_u/d \ell = 0$ at this order. The first equation \eqref{eq:RG_eqn_Z} generates an anomalous dimension for the order parameter field $\vec\phi$. The second equation \eqref{eq:RG_eqn_r} describes the flow of the tuning parameter $r$ and determines the correlation length exponent of the transition. Equations \eqref{eq:RG_eqn_c} and \eqref{eq:RG_eqn_cs} represent, respectively, the flow of the velocities of the $\vec \phi$ field and the longitudinal phonons. The elastic coupling $\lambda$ explicitly breaks Lorentz invariance triggering the flow of $c$ and $c_L$. Note that this can be also interpreted as a correction to the dynamical exponent $z \neq 1$ by demanding the vanishing of one of the corresponding $\beta$ functions. Choosing however $z=1$, both velocities will decrease under RG transformation. Finally, equations \eqref{eq:RG_eqn_u} and \eqref{eq:RG_eqn_l} describe the flow of the vertices. 

The RG flow of the four equations \eqref{eq:RG_eqn_c}, \eqref{eq:RG_eqn_cs}, \eqref{eq:RG_eqn_u} and \eqref{eq:RG_eqn_l} is coupled. It turns out that the flow can be simplified and the four coupled equations can be reduced to only three by introducing the dimensionless parameters
\begin{align}\label{eq:reparam_coupling}
    \vr &= \frac{c_L}{c},\qquad 
    \tu = \frac{u}{4!\, c^3},\qquad 
    \tl^2 =    \frac{\lambda^2}{2\rho c_L^2 c^3},
\end{align}
where $\vr$ is the ratio of the two velocities. Their RG flow is governed by the closed set of equations
\begin{align}
    \label{eq:RG_eqn_vr}
    \frac{d\vr}{d\ell} &=
        \frac{1}{4\pi^2} \left(\frac{16}{3} \frac{\vr^2}{(1+\vr)^3} - N\vr \right) \tl^2, \\
    \label{eq:RG_eqn_tu}
    \frac{d\tu}{d\ell} &= \epsilon \tu
        - \frac{N+8}{2\pi^2} \tu^2
        + \frac{6}{\pi^2} \frac{\vr(2 + \vr)}{(1+\vr)^2} \tu \tl^2 
       - \frac{2}{\pi^2} \frac{\vr(1 + 3\vr + \vr^2)}{(1+\vr)^3} \tl^4,
    \\
    \label{eq:RG_eqn_tl}
    \frac{d\tl^2}{d\ell} &= \epsilon \tl^2
        - \frac{N+2}{\pi^2} \tu \tl^2 + \frac{1}{2\pi^2} \left(N + 4 - \frac{4}{(1+\vr)^2}\right) \tl^4,
\end{align} 
\end{widetext}
where we have set $d = 3-\epsilon$. Note that, remarkably, these equations do not depend on the so far arbitrary dynamical exponent $z$.

For completeness, we also list the RG equations
 for $Z_\phi$, the tuning parameter $r$ and the velocities
in terms of dimensionless variables,
\begin{align}
\frac{d \log Z_\phi}{d \ell} & =- \frac{2}{\pi^2} \frac{\vr}{(1 + \vr)^3} \tl^2
 \\
 \label{eq:RG_eqn_tr}
\frac{d\log r}{d\ell} &= 2z
        - \frac{N+2}{2\pi^2} \tu
        + \frac{1}{\pi^2} \frac{\vr^2(3 + \vr)}{(1+\vr)^3} \tl^2,
   \\
       \frac{d \log c^2}{d \ell}
        &= 2z - 2 - \frac{8}{3\pi^2} \frac{\vr}{(1+\vr)^3} \tl^2,
\\ \label{eq:RG_eqn_csr}
\frac{d \log c_L^2}{d \ell}
        &= 2z - 2 - \frac{N}{2\pi^2} \tl^2.
    \end{align}

\subsection{Renormalization group flow and fixed points}
\label{subsec:RGFlow}

\begin{table*}
\begin{center} 
\def\arraystretch{2.0}
\setlength{\tabcolsep}{0.8em}
\begin{tabular}{|c|c|c|c|c|c|}
    \hline
    Fixed point & $\vr$ & $\tu$ &{$\tl^{2}$} & $z_{\rm Phonon}$ & %$\nu$ & 
    $\frac{d\log |\tl|}{d\ell}|_{\tl^2 = 0}$ \\ 
    \hline
    G& $\mathbb{R}_{>0}$ & 0 & 0
        %& 2 & $\frac{1}{2}$
        & 1 %&$\frac{1}{2}$ 
        & $\frac{\epsilon}{2}$\\
    \hline
    WF & $\mathbb{R}_{>0}$ & $\frac{2 \pi^2}{N+8}\epsilon$ & 0
        & $1$ %&$\frac{1}{2} + \frac{N+2}{4(N+8)}\epsilon$
        & $\frac{1}{2}\frac{4-N}{N+8}\epsilon$\\
      \hline
%       G$^* & 0 & 0 & $- \frac{2\pi^2}{N} \epsilon$
%        %& $2 - \frac{(N+2)\epsilon}{(N+8)}$ & $\frac{1}{2} + \frac{(N+2)\epsilon}{4(N+8)}$
%        & $1 - \frac{1}{2}\epsilon$ %&  $\frac{1}{2} + \frac{1}{4}\epsilon$
%        & n/a %$\frac{4-N}{2(N+8)}\epsilon$
%        \\        
%    \hline
    WF$^* (N>4)$ & 0 & $\frac{2 \pi^2}{N+8}\epsilon$ & ${\frac{2\pi^2(N-4)}{N(N+8)}}\epsilon$
        %& $2 - \frac{(N+2)\epsilon}{(N+8)}$ & $\frac{1}{2} + \frac{(N+2)\epsilon}{4(N+8)}$
        & $1 + \frac{N-4}{2(N+8)}\epsilon$ %&  $\frac{1}{2} + \frac{N+2}{4(N+8)}\epsilon$
        & n/a %$\frac{4-N}{2(N+8)}\epsilon$
        \\
\hline
\end{tabular}
\end{center}
\caption{Three fixed points characterize the RG flow in $d=3-\epsilon$ spatial dimensions. At zero elastic coupling $\tl^2 = 0$, there exists the conventional Gaussian (G) and Wilson-Fisher (WF) fixed point of the $\phi^4$ theory with a vanishing or finite self-interaction $\tu$, respectively. At G and WF, the ratio of velocity  $\vr = c_L/c$ assumes a finite value, $\vr \in \mathbb{R}_{>0}$. For $\epsilon > 0$, the elastic coupling $\tl^2$ is relevant with respect to G and with respect to WF in case $N < 4$, see last column. An additional fixed point WF$^*$ then arises in the limit $\vr \to 0$ for which the elastic degrees of freedom can be characterized by a dynamical exponent $z_{\rm phonon} \neq 1$.  
}
\label{tbl:fixedpt}
\end{table*}

For a vanishing elastic coupling $\lambda = 0$, the equations  
reduce to the well-known RG equations of the $\phi^4$ theory with a Gaussian (G) and a Wilson-Fisher (WF) fixed point at $r = 0$, see Table \ref{tbl:fixedpt}.
Note that at both fixed points the dynamical exponent is $z=1$; this ensures that both velocities, $c_L$ and $c$, remain invariant under RG transformations.

The scaling dimension of the dimensionless elastic coupling at $\tl^2 = 0$ is
\begin{align}
\frac{d\log |\tl|}{d\ell}\Big|_{\tl^2 = 0} = \frac{\epsilon}{2} - \frac{N+2}{2\pi^2} \tu.  
\end{align}
The elastic coupling is relevant and non-perturbative if the right-hand side is larger than zero. This indeed coincides with the criterion given in Eq.~\eqref{Criterion}, $\alpha_q = 2-\nu (d+z) > 0$ as we demonstrate in the following. At $\tl^2 = 0$ the correlation length exponent reads $1/\nu = \frac{d\log r}{d\ell}|_{\tl^2 = 0}  = 2 -\frac{N+2}{2\pi^2} \tu$ and $z=1$. So we get up to corrections of order $\mathcal{O}(\tu^2, \tu \epsilon)$,
\begin{align}
\alpha_q|_{\tl^2 = 0} = \frac{\epsilon}{2} - \frac{N+2}{2\pi^2} \tu.
\end{align}
At the Gaussian fixed point, $\tu = 0$, and the elastic coupling is relevant for $\epsilon > 0$. At the WF fixed point $\tilde u = \frac{2\pi^2}{N+8} \epsilon$ and the elastic coupling is relevant for $N < 4$.

For finite $\lambda \neq 0$ and $N \geq 1$, the velocity ratio $\vr$ always flows to smaller values, and its RG equation Eq.~\eqref{eq:RG_eqn_vr} only vanishes for $\vr = 0$. As both velocities decrease under RG, it follows that the longitudinal phonon velocity $c_L$ decreases faster than $c$. The flow is towards an instability of the crystal as will be further discussed below. Interestingly, 
Eq.~\eqref{eq:RG_eqn_vr} also vanishes for a finite ratio $\vr$ provided that $N \leq \frac{64}{81} \approx 0.79$ potentially giving rise to a non-trivial fixed point. In the following, we focus however on $N \geq 1$.

In the limit $\vr \to 0$, an additional fixed point WF$^*$ arises in the set of RG equations (\ref{eq:RG_eqn_vr})-(\ref{eq:RG_eqn_tl}) with a finite value for the dimensionless elastic coupling $\tl^2$,  see Table \ref{tbl:fixedpt}. This new fixed point is only present in case that $\tl^2$ is irrelevant with respect to the WF fixed point. If $\tl^2$ however exceeds a threshold value, it enables a flow towards strong coupling, as will be discussed in more detail below. Interestingly, WF$^*$ can be characterized by a renormalized dynamical exponent for the phonon degrees of freedom, $z_{\rm phonon} \neq 1$, by demanding that the scaling dimension of the longitudinal phonon velocity vanishes, see Eq.~\eqref{eq:RG_eqn_csr}.

The direction of the RG flow within the $(\tu, \tl^2)$ plane is determined by the matrix
\begin{align} \label{ScalingVariables}
&\left(\begin{array}{cc}
\frac{\partial}{\partial \tu} \frac{d\tu}{d\ell} & \frac{\partial}{\partial \tl^2} \frac{d\tu}{d\ell} \\
\frac{\partial}{\partial \tu} \frac{d\tl^2}{d\ell} & \frac{\partial}{\partial \tl^2} \frac{d\tl^2}{d\ell}
\end{array}\right)
\\\nonumber &
\overset{\vr \to 0}{=}
\left(\begin{array}{cc}
\epsilon - \frac{N+8}{\pi^2} \tu & 0  \\
\\ -\frac{N+2}{\pi^2} \tl^2 & \epsilon + \frac{1}{\pi^2} (N \tl^2 - (N+2) \tu)
\end{array}\right).
\end{align}
In the limit $\vr \to 0$, the flow is along the two directions $(0,1)$ and $(6\tu + N\tl^2 ,(N+2) \tl^2)$ with eigenvalues $\epsilon + \frac{1}{\pi^2} (N \tl^2 - (N+2) \tu)$ and $\epsilon - \frac{N+8}{\pi^2} \tu$, respectively. Using the respective fixed point values for $\tu$ and $\tl^2$, this specifies also the scaling fields and scaling dimensions of each fixed point. In the following, we will discuss in more detail the RG flow for various representative examples.

\subsubsection{RG flow for d=2.99 dimensions}

\begin{figure}
    \centering
    \includegraphics[width=0.8\columnwidth]{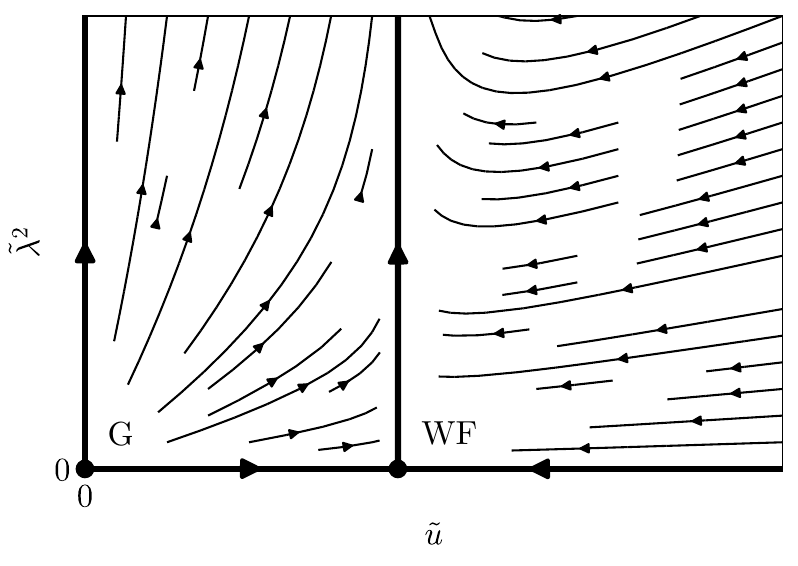}
    \caption{RG flow for $d = 2.99$ dimensions $(\epsilon = 0.01)$, $\vr \ll 1$ and $1\leq N \leq 4$. Both, the G and WF fixed point are unstable with respect to $\tl^2$ resulting in runaway flow. For $N = 4$, the elastic coupling is only marginally relevant with respect to WF.}
    \label{fig:rg_flow_N_below_4}
\end{figure}

\begin{figure}
    \centering
    \includegraphics[width=0.8\columnwidth]{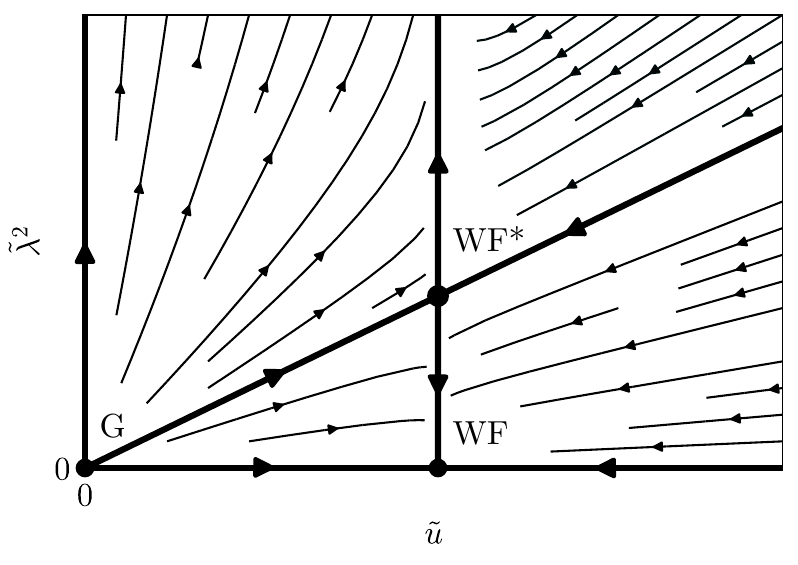}
    \caption{RG flow for $d = 2.99$ dimensions $(\epsilon = 0.01)$, $\vr \ll 1$ and $N > 4$. The G and WF fixed points are unstable and stable, respectively. The additional fixed point WF$^*$ is repulsive with respect to $\tl^2$ leading to runaway flow for sufficiently large elastic coupling.}
    \label{fig:rg_flow_N_above_4}
\end{figure}

Below three spatial dimensions, $\epsilon > 0$, the Gaussian fixed point is unstable  with respect to both $\tu$ and $\tl^2$. The flow for $\tl^2 = 0$ is towards the stable WF fixed point, but for finite $\tl^2>0$ it depends on the number of components $N$. For $1 \leq N < 4$, on the one hand, the WF fixed point is unstable and the flow is always towards strong coupling, see Fig.~\ref{fig:rg_flow_N_below_4}. For $N > 4$, on the other hand, the WF fixed is perturbatively stable, see Fig.~\ref{fig:rg_flow_N_above_4}. In this case, however, the repulsive fixed point WF$^*$ emerges such that the flow is towards strong coupling for sufficiently large $\tl^2$. The separatrix within the $(\tu ,\tl^2)$ plane, that separates weak from strong coupling flow, possesses the slope $(N-4)/N$ for $\vr \to 0$, which can be obtained from the scaling variables of WF$^*$, see Eq.~\eqref{ScalingVariables}. For $N=4$ this slope vanishes and the two fixed points WF and WF$^*$ merge while $\tl^2$ remains relevant albeit only marginally.

\begin{figure}
    \centering
    \includegraphics[width=0.8\columnwidth]{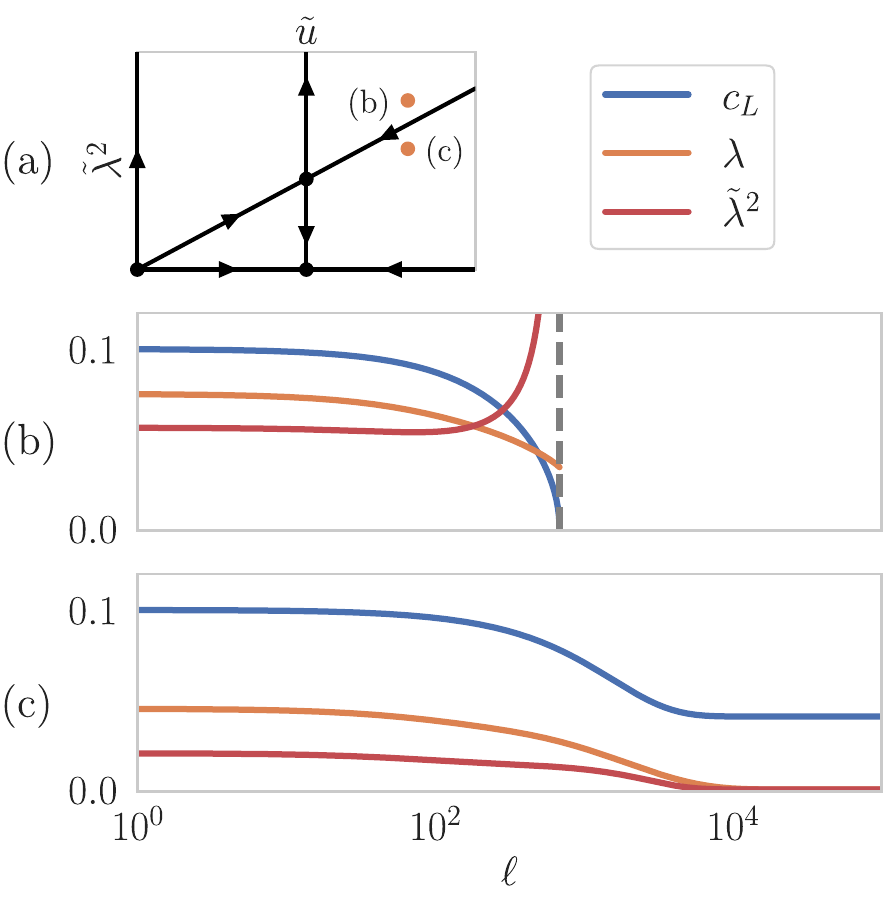}
    \caption{
    RG flow for $d=2.99$ and $N=5$ of the phonon velocity $c_L$, the elastic coupling $\lambda$, and the dimensionless elastic coupling $\tl^2$ using starting values $c_0=1$, $c_{0,L}=0.1$ and $u_0 = 0.5$. 
    The starting value for panel (b) $\lambda_0 = 0.015$ is located above the separatrix as indicated in (a), leading to runaway flow: the phonon velocity vanishes at some finite RG scale $\ell$. The starting value for panel (c) $\lambda_0 = 0.009$ is located below the separatrix, and the flow is towards the WF fixed point. Here, the velocity $c_L$ saturates at a finite value.
}
    \label{fig:numerical_example}
\end{figure}

In the following, we will further elaborate on the RG flow for $N>4$ where both fixed points, WF and WF$^*$, are present. In order to elucidate the physics of the runaway flow, we present in Fig.~\ref{fig:numerical_example} the flow of the dimensionful variables $c_L$ and $\lambda$ as well as the dimensionless coupling $\tl^2$ for starting values below and above the separatrix. 
The flow of the longitudinal phonon velocity $c_L$ is always towards smaller values. 
For starting values above the separatrix, the flow of $\tl^2$ is towards strong coupling and the phonon velocity $c_L$ vanishes at some finite RG scale $\ell$, see panel (b). For starting values within the basin of attraction of the WF fixed point, see panel (c), the dimensionless elastic coupling $\tl^2$ flows sufficiently fast to zero such that the flow for $c_L$ stops and the phonon velocity remains finite at lowest energies. Note however that this does not automatically imply that the crystal remains macroscopically stable, see section \ref{sec:CrystalStability}.
 
\begin{figure}
    \centering
    \includegraphics[width=0.8\columnwidth]{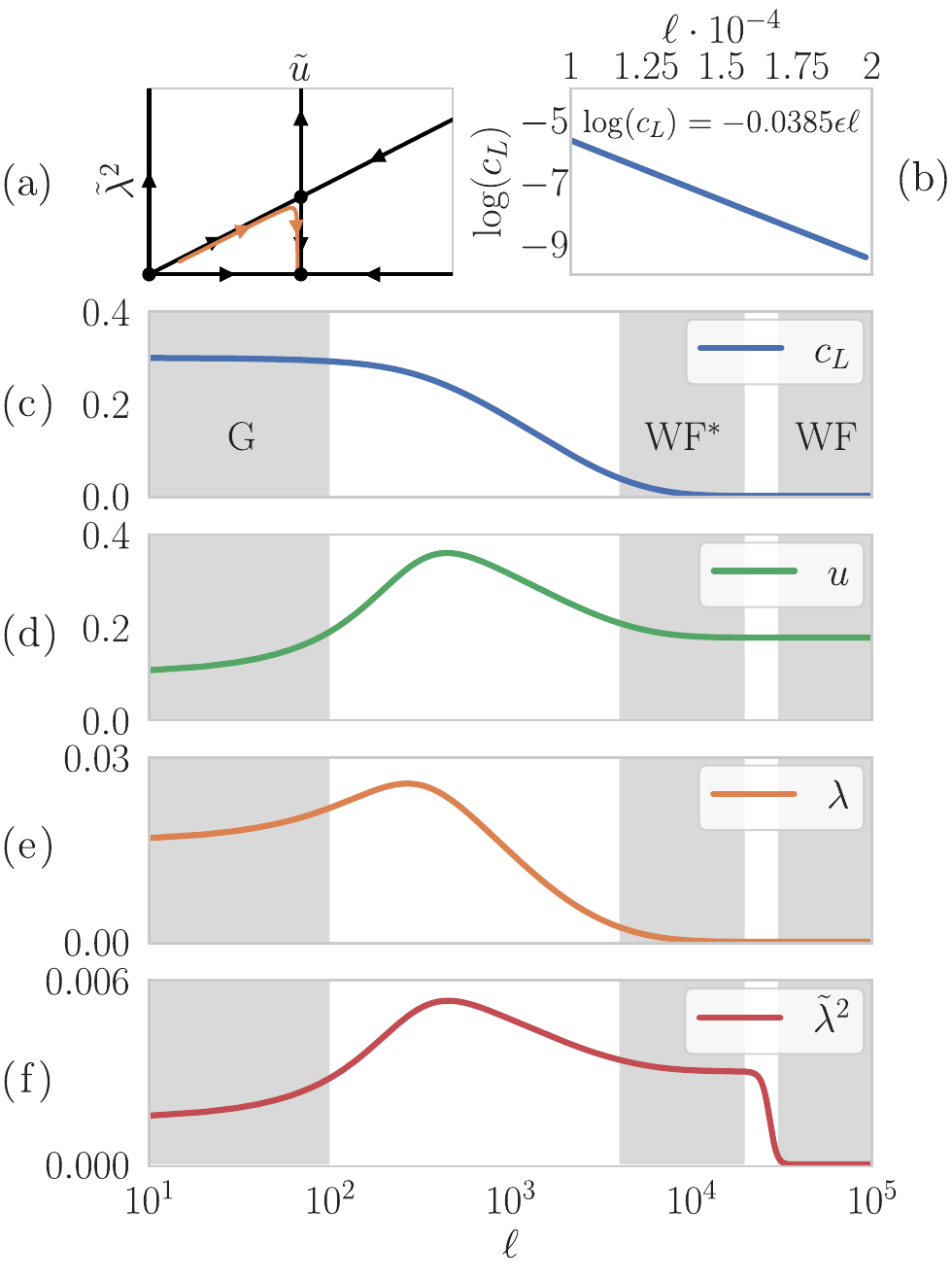}
    \caption{RG flow for $d=2.99$ and $N=5$ illustrating the influence of the WF$^*$ fixed point. Starting values just below the separatrix are chosen, $c_0=1$, $c_{L,0}=0.3$, $u_0 = 0.1$ and $\lambda_0 \approx0.0163$. The resulting flow within the $(\tu, \tl^2)$ plane is shown in panel (a). Panels (c)-(f) display the phonon velocity $c_L$, the dimensionful interactions $u$ and $\lambda$, as well as the dimensionless elastic coupling $\tl^2$ as a function of the RG scale $\ell$, respectively.
    The shaded region indicate the range of $\ell$ governed by a specific fixed point whereas the white regions correspond to crossovers. 
    Initially, the flow is still influenced by the Gaussian fixed point G. There is an extended range of $\ell$ where the flow is dominated by WF$^*$ leading to a power-law dependence of the phonon velocity with the expected  exponent  $\delta z = \frac12 \frac{N-4}{N+8} \epsilon \approx 3.846 \times10^{-4}$, as shown in more detail in panel (b). Eventually, the flow is governed by the WF fixed point when $\tl^2$ drops quickly to zero. }     
\label{fig:regimes}
\end{figure}

The influence of the WF$^*$ fixed point materializes if the initial values are located on or very close to the separatrix. In this case, the phonon velocity eventually decreases as $c_L(\ell) \sim e^{-\delta z\,  \ell}$ that can be interpreted as a correction to the dynamical exponent of the phonons $z_{\rm phonon} = 1 + \delta z$, see Table \ref{tbl:fixedpt}, where $\delta z = \frac12 \frac{N-4}{N+8} \epsilon$. This is illustrated in Fig.~\ref{fig:regimes}.
 
\subsubsection{RG flow for d=3 dimensions}

\begin{figure}
    \centering
    \includegraphics[width=0.8\columnwidth]{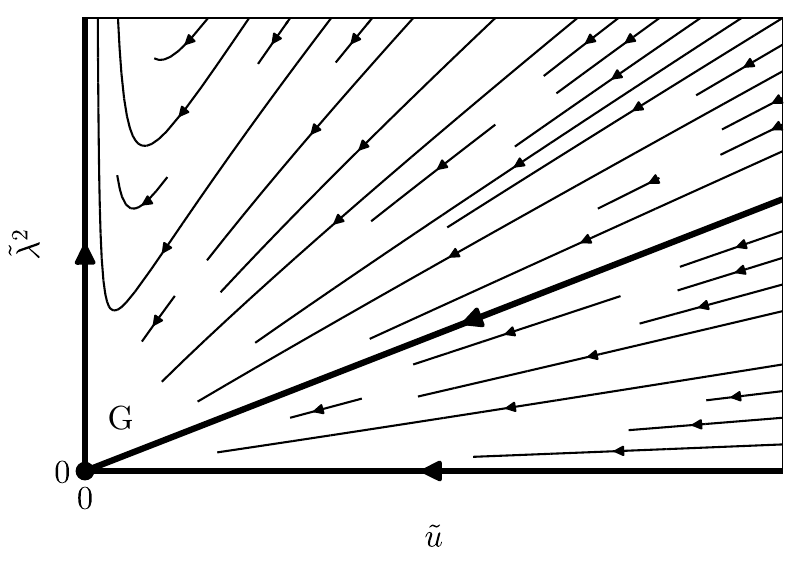}
    \caption{RG flow at the upper critical dimension $d = 3$ for $\vr \ll 1$ and $N>4$.
    There exists a separatrix below which the flow is towards the Gaussian fixed point. Above the separatrix the flow is towards strong coupling.}
    \label{fig:rg_flow_N_above_4_at_critical_dimension}
\end{figure}

\begin{figure}
    \centering
    \includegraphics[width=0.8\columnwidth]{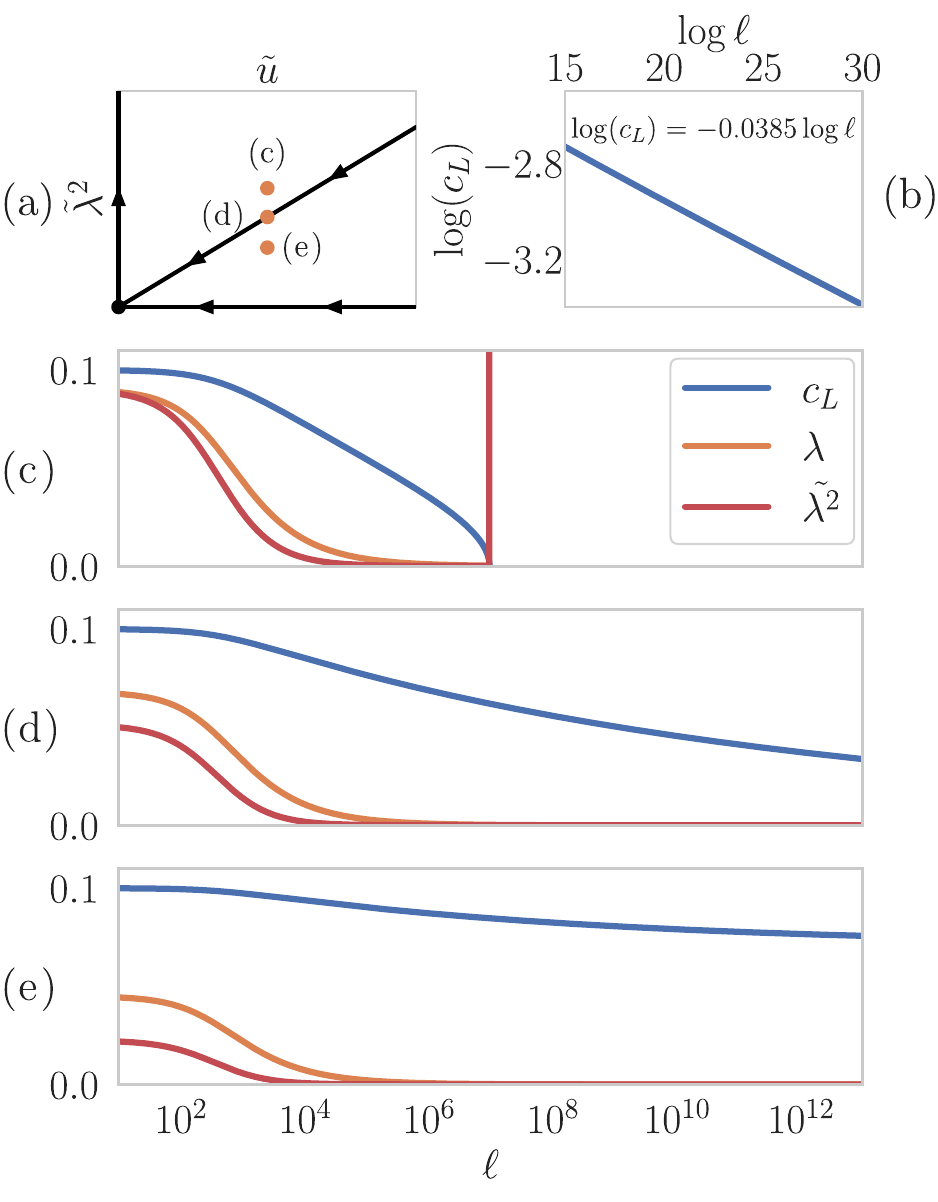}
    \caption{RG flow for $d=3$ and $N=5$ 
   of the phonon velocity $c_L$, the elastic coupling $\lambda$, and the dimensionless elastic coupling $\tl^2$ using starting values $c_0=1$, $c_{L,0}=0.1$, $u_0 = 0.1$. Panel (a) illustrates the choice for the starting value of $\lambda$. For panel (c)-(d) the starting values 
    $\lambda_0 = 0.006$, $\lambda_0 \approx 0.0045$, and $\lambda_0 = 0.003$ are chosen that are located above, on, and below the separatrix, respectively. Above the separatrix (c) the phonon velocity vanishes at a finite RG scale. On the separatrix (d) the phonon velocity $c_L \sim \ell^{-\#}$ vanishes logarithmically  with exponent $\# = \frac{N-4}{2(N+8)} = 0.03846..$, as illustrated in panel (b). Below the separatrix (e) $c_L$ saturates in the large $\ell$ limit.}
    \label{fig:10}
\end{figure}

At the upper critical dimension $d=3$ both couplings $\tu$ and $\tl^2$ are marginal and the RG flows only logarithmically. In general, the solution of the three coupled equations \eqref{eq:RG_eqn_vr} - \eqref{eq:RG_eqn_tl} is involved but it simplifies in the limit of small $\vr \ll 1$. The flow within the $(\tu, \tl^2)$ plane is then asymptotically decribed by 
\begin{align}
%    \frac{d\vr}{d\ell} &=-\frac{N\vr\tl}{4\pi^2} \\
    \frac{d\tu}{d\ell} &\approx -\frac{(N+8) }{2\pi^2} \tu^2, \\
    \frac{d\tl^2}{d\ell} &\approx -\frac{(N+2)}{\pi^2} \tu\tl^2 + \frac{N}{2\pi^2} \tl^4.
\end{align}
The first equation is independent of $\tl^2$, and it possesses the solution
\begin{equation}
    \tu(\ell) = \frac{\tu_0}{a\tu_0 \ell + 1} ,
\end{equation}
where $\tu(0) = \tu_0$ and we abbreviated $a = \frac{(N+8)}{2\pi^2}$. This implies that 
$\tu$, on the one hand, is always marginally irrelevant with the asymptotic behavior $\tu \sim \frac{2\pi^2}{(N+8) \ell}$. The Gaussian fixed point, on the other hand, is always unstable with respect to the elastic coupling. Solving the second equation at $\tu =0$ one finds $\tl^2(\ell) = \tl^2_0/(1-\frac{\tl^2_0 N \ell}{2\pi^2})$ with a pole at the RG scale $\ell = 2\pi^2/(\tl^2_0 N)$ with the initial value $\tl^2(0) = \tl^2_0$. The dimensionless elastic coupling thus reaches infinity at a finite RG scale $\ell$ indicating that the phonon velocity $c_L$ vanishes at a finite scale $\ell$.

The behaviour  within the $(\tu, \tl^2)$ plane away from the two axis depends on the number of components $N$ reminiscent of the behaviour for finite $\epsilon$. The general solution reads
\begin{equation} \label{Flowd3}
    \tl^2(\ell) = \begin{cases}
        \frac{\frac{N-4}{N} \tu_0}{
            \left(\frac{N-4}{N} \frac{\tu_0}{\tl_0^2} - 1\right)
            \left(a \tu_0 \ell + 1\right)^{\frac{2 (N+2)}{N+8}}
            + a \tu_0 \ell + 1
        } & \text{for}~N \neq 4 \\
        \frac{3 \tu_0 \tl_0^2}{\left(a \tu_0 \ell+1\right) \left(3 \tu_0-\tl_0^2
           \log \left(a \tu_0 \ell+1\right)\right)} & \text{for}~N = 4 .
    \end{cases}
\end{equation}
For $1 \leq N \leq 4$, we find that the elastic coupling $\tl^2$ diverges at a finite RG scale implying that the axis $\tl^2 = 0$ is unstable with respect to a small elastic coupling. For $N>4$ the flow is only towards strong coupling provided that the initial value of the elastic coupling exceeds the threshold value,
\begin{equation}
    \tl_0^2 > \frac{N-4}{N} \tu_0,
\end{equation}
which also defines the separatrix separating the flow to strong and to weak coupling. This is illustrated in Figs.~\ref{fig:rg_flow_N_above_4_at_critical_dimension} and \ref{fig:10}.

\subsection{Microscopic and macroscopic instability of the crystal lattice}
\label{sec:CrystalStability}

The RG flow discussed in the last section characterizes the microscopic fluctuations of the $\vec \phi$ field as well as the longitudinal acoustic phonons within the bulk of the system. 
When the flow of the dimensionless elastic coupling is towards strong coupling, $\tl^2 \to \infty$, we found that the longitudinal phonon velocity, $c_L$, vanishes at a {\it finite} RG scale, see for example Fig.~\ref{fig:numerical_example}(b) and Fig.~\ref{fig:10}(c). This signals a microscopic elastic instability within the bulk of the crystal. Such an instability will generically trigger a first-order isostructural transition irrespective of the boundary conditions.

If the RG flow of $\tl^2$ is either towards weak coupling $\tl^2 \to 0$ or towards the fixed point value of WF$^*$, 
the longitudinal phonon velocity $c_L$, respectively, remains finite at all scales, see Fig.~\ref{fig:numerical_example}(c) and Fig.~\ref{fig:10}(e), or it vanishes only asymptotically for $\ell \to \infty$, see Fig.~\ref{fig:regimes} and Fig.~\ref{fig:10}(d). This implies that at any finite RG scale $\ell$ the microscopic phonon degrees of freedom are stable. 

However, this does not imply that the crystal also remains stable macroscopically. The macroscopic elastic stability of a crystal is generally determined by the elastic moduli, which are required to be positive.
If the bulk modulus $K$ vanishes for free boundary conditions, i.e., at constant hydrostatic pressure $P$, the crystal becomes unstable. This isostructural instability generically results in a first-order transition because the Landau potential for the bulk strain possesses a cubic term. For vanishing $K$ and isotropic elasticity, however, the crystal might still be stabilized by pinned boundary conditions as specified by Bergmann and Halperin \cite{BergmanHalperin1976}. 

For isotropic elasticity, the longitudinal and transversal sound velocity, see Eq.~\eqref{PhononPropagator}, are given in terms of the bulk and shear moduli by 
\begin{align} \label{PhononVelocities}
c^2_L = \frac{1}{\rho} \left(K + \frac{4}{3}\mu\right), \quad
c^2_T = \frac{1}{\rho} \mu.
\end{align}
Importantly, at a macroscopic instability where the bulk modulus $K$ vanishes and eventually turns negative both phonon velocities remain finite. As a result, the isostructural transition of a crystal is a genuine mean field transition without critical phonon fluctuations\cite{Cowley1976}. 
As the transversal phonon modes do not couple to the $\vec\phi$ critical degrees of freedom, $c_T$ is invariant under RG transformation and so is the shear modulus $\mu$. The RG flow of the bulk modulus $K$ is therefore determined by the one of the longitudinal velocity 
\begin{align}
K(\ell) = \rho \left(c_L^2(\ell) - \frac{3}{4} c_T^2 \right).
\end{align}
Note, in particular, that this holds because we obtained at one-loop order $Z_u = 1$ for the wavefunction renormalization of the phonons. As $c_L$ decreases under RG the bulk modulus vanishes at an RG scale $\ell^*$ where $c_L^2(\ell^*) = \frac{3}{4} c_T^2$. Due to the finite shear modulus, the macroscopic elastic instability of the crystal, in general, preempts the microscopic instability.

\begin{figure}
    \centering
    \includegraphics[width=0.1\columnwidth]{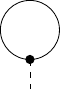}
    \caption{Diagram with a closed loop of the $\vec\phi$ propagator (solid line) that generates an internal hydrostatic pressure.}
    \label{fig:ballon}
\end{figure}

When the dimensionless elastic coupling $\tl^2$ flows to strong coupling, there exists always a finite scale $\ell^*$ where $K(\ell^*) = 0$ and the crystal becomes macroscopically unstable at constant $P$ before the microscopic instability develops. However, such a macroscopic instability can also exist even when the flow of $\tl^2$ is towards weak coupling. In this case, $c_L$ decreases during the RG flow as long as $\tl^2$
is finite but $c_L$ eventually saturates at a finite value, see Fig.~\ref{fig:numerical_example}(c) and Fig.~\ref{fig:10}(e). If this saturation value implies a negative bulk modulus $K(\ell) < 0$, there exist even in such a case a critical RG scale $\ell^*$ where the macroscopic instability develops. 
We conclude that a macroscopic elastic instability can occur even when the microscopic elastic degrees of freedom remain non-critical and the flow of $\tl^2$ is towards weak coupling. 

A special situation arises when the elastic coupling flows towards the WF$^*$ fixed point. In this case, the longitudinal phonon velocity 
$c_L(\ell) \sim e^{-\delta z\, \ell}$
vanishes as a function of increasing correlation length $\xi = \xi_0 e^\ell$ with a power law
\begin{align} \label{PhononVelocityPowerLaw}
c_L &\sim \xi^{-\delta z}
\end{align}
with $\delta z = z_{\rm phonon} - 1  = \frac{N-4}{2(N+8)} > 0$, see Table \ref{tbl:fixedpt}. In this case, a critical RG scale $\ell^*$ also exist where the bulk modulus vanishes. For free boundary conditions at constant hydrostatic pressure $P$, the vanishing of the bulk modulus will trigger a first-order transition thus preempting the flow towards WF$^*$. However, for pinned boundary conditions the first-order transition might be avoided such that the asymptotic properties of the new fixed points become accessible.

At the upper critical dimension $d=3$, the flow is always towards strong coupling for $N \leq 4$. For $N>4$ it depends on the initial conditions, and the flow is only towards strong coupling for sufficiently large values of $\tl^2$. The weak coupling flow on the separatrix is governed by the asymptotic behavior, see Eq.~\eqref{Flowd3},
\begin{align}
\tl^2(\ell) &\approx  \frac{2\pi^2(N-4)}{N(N+8)} \frac{1}{\ell}.
\end{align}
Plugging this into the RG equation for the phonon velocity Eq.~\eqref{eq:RG_eqn_csr}, we obtain 
\begin{align}
c_L \sim \left[\log (\xi/\xi_0)\right]^{- \frac{N-4}{2(N+8)}},
\end{align}
as a function of $\xi = \xi_0 e^\ell$. On the separatrix the phonon velocity thus vanishes logarithmically reminiscent of the behavior in 
Eq.~\eqref{PhononVelocityPowerLaw}, see Fig.~\ref{fig:10}. Below the separatrix the flow of $c_L$ is found to saturate at a finite value. Whether an elastic first-order transition develops for the weak coupling flow again depends on the value of the shear modulus and the boundary conditions.

The $\vec \phi$ degrees of freedom also generate an internal hydrostatic pressure on the bulk strain. This pressure is proportional to the closed loop of the $\vec \phi$ propagator, see Fig.~\ref{fig:ballon}. The resulting response of the system also 
depends on the boundary conditions. For pinned boundary conditions, this internal pressure is compensated by the external forces imposing the boundary conditions. For free boundary conditions at constant $P$, the macroscopic bulk strain $E$ will respond to the internal pressure. 
In this manner, the critical $\vec \phi$ degrees of freedom cause a non-analytic dependence of the 
expansivity $E(r,T)$ on the tuning parameter $r$ and the temperature $T$.
This results in a bulk thermal expansion $\partial_T E$ with characteristic quantum critical signatures \cite{Zhu2003,Garst2005}. However, if the bulk modulus becomes small and the system is close to the isostructural instability, Hooke's law will break down and the elastic response will be non-linear.

 \section{Summary \& Discussion}
 \label{sec:Discussion}
 
A stability analysis of a second-order quantum critical point with respect to an elastic coupling predicts a breakdown of perturbation theory if the criterion of Eq.~\eqref{Criterion} is fulfilled, i.e., $\alpha_q = 2-\nu (d+z) > 0$. In the present work, we confirmed this explicitly for the $\phi^4$ theory with $O(N)$ symmetry and dynamical exponent $z = 1$ close to spatial dimensions $d=3-\epsilon$ assuming isotropic elasticity. 
The Wilson-Fisher fixed point (WF) possesses the exponent $\alpha_q = \epsilon \frac{4-N}{2(N+8)}$ and, consistent with expectations, it is unstable with respect to an elastic coupling for $1 \leq  N < 4$. We found that this coupling flows under RG transformation towards strong coupling, $\tl^2 \to \infty$, see Fig.~\ref{fig:rg_flow_N_below_4}.
In this case, the longitudinal phonon velocity vanishes at a finite RG scale triggering a microscopic elastic instability. As a result, the crystal becomes microscopically and macroscopically unstable resulting in a first-order isostructural transition for any boundary conditions, in particular, for both constant hydrostatic pressure $P$ and constant volume $V$. 
For $N= 4$ we find that the elastic coupling remains relevant but only marginally.

For $N >4$, the exponent $\alpha_q < 0$ and the impact of an elastic coupling is expected to be perturbative. Consistent with this expectation, the RG flow for small $\tl^2$ is found to be towards weak coupling, $\tl^2 \to 0$, see  Fig.~\ref{fig:rg_flow_N_above_4}. However, if $\tl^2$ exceeds a threshold value the situation turns out be qualitatively different. The presence of a repulsive fixed point WF$^*$ gives rise to a separatrix beyond which the flow of $\tl^2$ is again towards strong coupling, that again induces a first-order transition for any boundary conditions.
On the separatrix the flow is towards the fixed point WF$^*$, and the dimensionless elastic coupling flows towards a constant value. As a consequence, the longitudinal phonon velocity will decrease as a power law with increasing correlation length, $c_L \sim \xi^{-\delta z}$,  where the exponent $\delta z = z_{\rm Phonon} - 1$ can be interpreted as a correction to the dynamical exponent of the phonons, see Table \ref{tbl:fixedpt}. In this case, the velocity $c_L$ vanishes only asymptotically such that for any finite $\xi$ the crystal remains microscopically stable. However, the bulk modulus vanishes at a finite value of $\xi$ and whether the crystal remains macroscopically stable depends on the boundary conditions. For free boundary conditions, i.e., at constant hydrostatic pressure $P$ a first-order transition is expected, while for pinned boundary conditions the macroscopic elastic instability might be avoided stabilizing the criticality of the new fixed point WF$^*$.

If the flow of $\tl^2$ is towards weak coupling, the phonon velocity $c_L$ saturates at a diminished but finite value, and microscopically the crystal remains stable. 
Depending on the shear modulus, a macroscopic instability can still develop in case that the bulk modulus turns negative, see Eqs.~\eqref{PhononVelocities}. Similarly as before, it depends on the boundary conditions whether the system then undergoes a first-order isostructural transition or not.

At the upper critical dimension $d=3$ all fixed points, G, WF, and WF$^*$ merge. Here, the flow of the elastic coupling $\tl^2$ for $1 \leq N \leq 4$ is towards strong coupling inducing an elastic first-order transition irrespective of the boundary conditions. 
For $N > 4$ a separatrix still separates the flow to strong and weak coupling, see Fig.~\ref{fig:rg_flow_N_above_4_at_critical_dimension}, reminiscent of the case for $\epsilon > 0$. We find that the flow of the longitudinal phonon velocity on the separatrix now vanishes logarithmically $c_L \sim [\log(\xi/\xi_0)]^{-\frac{N-4}{2(N+8)}}$, whereas below the separatrix the flow of $c_L$ saturates at a finite value. Here, the macroscopic stability of the crystal again depends on the boundary conditions.

\begin{figure}[t]
\includegraphics[width=0.8\columnwidth]{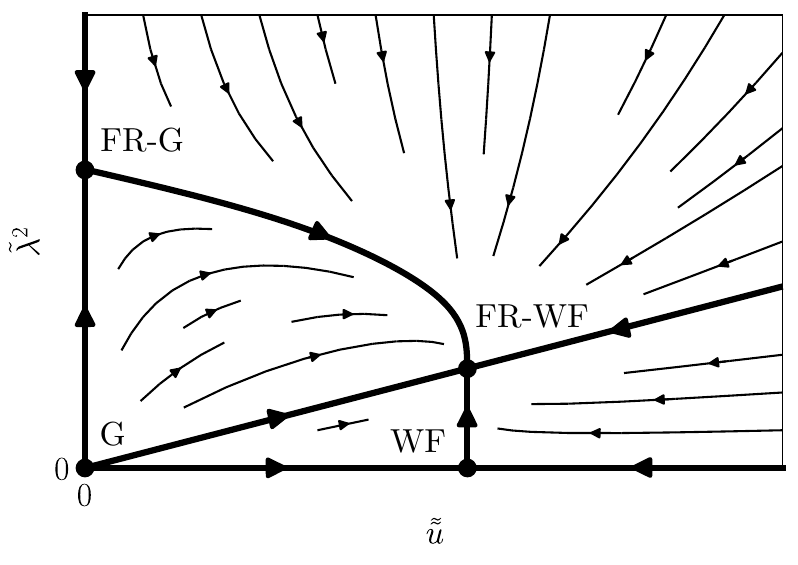}
\caption{
RG flow of the classical problem\cite{BergmanHalperin1976} for $d=4-\epsilon$ with $\epsilon = 0.01$ and $1 \leq N < 4$. There are four fixed points: Gaussian (G), Fisher-renormalized Gaussian (FR-G), Wilson-Fisher (WF) and Fisher-renormalized Wilson-Fisher (FR-WF). The FR-WF fixed point is stable within the $(\tilde \tu,\tl^2)$ plane, where $\tilde \tu = \tu - \tl^2$, and there is no runaway flow. For $N \geq 4$, the WF fixed point is instead stable with respect to the elastic coupling $\tl^2$ (not shown). 
} 
\label{fig:12}
\end{figure}

The sensitivity of the second-order quantum critical point with respect to the elastic coupling, that we find, is thus rather distinct from the classical case. In the classical problem for $d = 4-\epsilon$, runaway flow towards strong elastic coupling never occurs because there always exist stable fixed points\cite{BergmanHalperin1976}. Depending on the number of components $N$ either the WF or the Fisher-renormalized WF fixed point is stable for $\epsilon > 0$. It is interesting to compare the RG equations for the dimensionless couplings Eqs.~\eqref{eq:RG_eqn_vr} - \eqref{eq:RG_eqn_tl} of the quantum problem to the classical case. As shown in Ref.~\onlinecite{BergmanHalperin1976}, in the classical problem only the RG equations for the dimensionless parameters $\tu$ and $\tl^2$ are coupled. If the theory is restricted to the zero Matsubara modes, the integrals of Eqs.~\eqref{Integral1} and \eqref{Integral2} instead read
\begin{align}
I_1 &= \frac{1}{\beta} \int \frac{d\vec q}{(2\pi)^{d}}  
 \frac{1}{ c^2q^2 + r},\quad I_2(\vec q', \omega') = \frac{1}{\rho c_L^2} I_1,
\end{align}
with the inverse temperature $\beta$. Extracting their infrared divergences,   
the RG equations for the same dimensionless couplings \eqref{eq:reparam_coupling} but in $d = 4-\epsilon$ dimensions are obtained\cite{BergmanHalperin1976}
\begin{align} 
\label{ClassicalRGu}
\frac{\partial \tu}{\partial \ell} &= \epsilon \tu - \frac{N+8}{2 \pi^2 \beta c} \tu^2 + \frac{6}{\pi^2 \beta c} \tu \tl^2 - \frac{2}{\pi^2 \beta c} \tl^4,
\\
\label{ClassicalRGlambda}
\frac{\partial \tl^2}{\partial \ell} &= \epsilon \tl^2 - \frac{N+2}{\pi^2 \beta c} \tu \tl^2 + \frac{N+4}{2 \pi^2 \beta c} \tl^4.
\end{align}
They can be decoupled, however, after introducing $\tilde \tu = \tu - \tl^2$, that amounts to a reduced local self-interaction caused by the elastic degrees of freedom. The RG equation for the shifted $\tilde \tu$ then assumes the same form as in the absence of an elastic coupling such that $\tilde \tu$ converges to its value at the WF fixed point for $\epsilon > 0$, provided that the theory is stable $\tilde \tu > 0$. The remaining RG equation for $\tl^2$ then gives rise to two solutions such that in total four fixed points are obtained: G, WF, FR-G, and FR-WF, see Fig.~\ref{fig:12}.

In the quantum case, the velocities of both the critical $\vec \phi$ and phonon degrees of freedom, $c$ and $c_L$, flow under RG because the elastic coupling explicitly breaks Lorentz invariance. Their ratio $\vr = c_L/c$ also flows and influences the RG trajectories of both dimensionless couplings $\tu$ and $\tl^2$ signalling that space and time are intertwined at the quantum critical point. 
Interestingly, if we consider the limit $\vr \to \infty$ in the RG equations for $\tu$ and $\tl^2$, Eqs.~\eqref{eq:RG_eqn_tu} and \eqref{eq:RG_eqn_tl}, they acquire the same form as in the classical problem. 
This is consistent with the result of Ref.~\onlinecite{Chandra2020} that suggests that in the limit of vanishing mass density $\rho \to 0$, i.e., $\vr \to \infty$, the quantum problem in $d=3-\epsilon$ possesses the same fixed point structure as Eqs.~\eqref{ClassicalRGu} and \eqref{ClassicalRGlambda}. 
The fact that we found distinct behavior in the quantum theory can thus be traced to the RG flow of $\vr$ that does not flow to infinity but rather towards zero under RG transformations for $N \geq 1$. Note that the flow $\vr \to 0$ is characteristic for the quantum critical point considered in this work; for other quantum phase transitions the ratio of velocities could exhibit different scaling behavior, see for example Ref.~\onlinecite{Sitte2009}.

In this study, we limited ourselves to isotropic elasticity. In the classical case, it was found that anisotropic elasticity increases the tendency towards a microscopic elastic instability and thus towards a first-order isostructural transition\cite{DeMoura1976,BergmanHalperin1976,Nattermann1977}. 
We leave it for a future study whether the same holds true in the quantum case. 
With the caveat of isotropic elasticity,  our results directly apply to certain dimerized antiferromagnets like TlCuCl$_3$ \cite{Merchant2014} that exhibit a quantum phase transition in the $O(3)$ universality class and $z=1$. We predict that the latter
is unstable with respect to an elastic coupling. The quantum fluctuations will induce a microscopic elastic instability eventually leading to a isostructural first-order transition masking the putative quantum critical point in these materials.

\section*{acknowledgement}
	
Helpful discussions and a collaboration on a related topic with I. Paul are gratefully acknowledged. M.G. is  
supported by the Deutsche Forschungsgemeinschaft through TRR 288 - 422213477 (project A11).	
	
\bibliographystyle{apsrev4-1}
\bibliography{references}

\begin{thebibliography}{33}%
\makeatletter
\providecommand \@ifxundefined [1]{%
 \@ifx{#1\undefined}
}%
\providecommand \@ifnum [1]{%
 \ifnum #1\expandafter \@firstoftwo
 \else \expandafter \@secondoftwo
 \fi
}%
\providecommand \@ifx [1]{%
 \ifx #1\expandafter \@firstoftwo
 \else \expandafter \@secondoftwo
 \fi
}%
\providecommand \natexlab [1]{#1}%
\providecommand \enquote  [1]{``#1''}%
\providecommand \bibnamefont  [1]{#1}%
\providecommand \bibfnamefont [1]{#1}%
\providecommand \citenamefont [1]{#1}%
\providecommand \href@noop [0]{\@secondoftwo}%
\providecommand \href [0]{\begingroup \@sanitize@url \@href}%
\providecommand \@href[1]{\@@startlink{#1}\@@href}%
\providecommand \@@href[1]{\endgroup#1\@@endlink}%
\providecommand \@sanitize@url [0]{\catcode `\\12\catcode `\$12\catcode
  `\&12\catcode `\#12\catcode `\^12\catcode `\_12\catcode `\%12\relax}%
\providecommand \@@startlink[1]{}%
\providecommand \@@endlink[0]{}%
\providecommand \url  [0]{\begingroup\@sanitize@url \@url }%
\providecommand \@url [1]{\endgroup\@href {#1}{\urlprefix }}%
\providecommand \urlprefix  [0]{URL }%
\providecommand \Eprint [0]{\href }%
\providecommand \doibase [0]{http://dx.doi.org/}%
\providecommand \selectlanguage [0]{\@gobble}%
\providecommand \bibinfo  [0]{\@secondoftwo}%
\providecommand \bibfield  [0]{\@secondoftwo}%
\providecommand \translation [1]{[#1]}%
\providecommand \BibitemOpen [0]{}%
\providecommand \bibitemStop [0]{}%
\providecommand \bibitemNoStop [0]{.\EOS\space}%
\providecommand \EOS [0]{\spacefactor3000\relax}%
\providecommand \BibitemShut  [1]{\csname bibitem#1\endcsname}%
\let\auto@bib@innerbib\@empty
%</preamble>
\bibitem [{\citenamefont {Cowley}(1976)}]{Cowley1976}%
  \BibitemOpen
  \bibfield  {author} {\bibinfo {author} {\bibfnamefont {R.~A.}\ \bibnamefont
  {Cowley}},\ }\href {\doibase 10.1103/PhysRevB.13.4877} {\bibfield  {journal}
  {\bibinfo  {journal} {Phys. Rev. B}\ }\textbf {\bibinfo {volume} {13}},\
  \bibinfo {pages} {4877} (\bibinfo {year} {1976})}\BibitemShut {NoStop}%
\bibitem [{\citenamefont {Zacharias}\ \emph
  {et~al.}(2015{\natexlab{a}})\citenamefont {Zacharias}, \citenamefont {Paul},\
  and\ \citenamefont {Garst}}]{Zacharias2015}%
  \BibitemOpen
  \bibfield  {author} {\bibinfo {author} {\bibfnamefont {M.}~\bibnamefont
  {Zacharias}}, \bibinfo {author} {\bibfnamefont {I.}~\bibnamefont {Paul}}, \
  and\ \bibinfo {author} {\bibfnamefont {M.}~\bibnamefont {Garst}},\ }\href
  {\doibase 10.1103/PhysRevLett.115.025703} {\bibfield  {journal} {\bibinfo
  {journal} {Phys. Rev. Lett.}\ }\textbf {\bibinfo {volume} {115}},\ \bibinfo
  {pages} {025703} (\bibinfo {year} {2015}{\natexlab{a}})}\BibitemShut
  {NoStop}%
\bibitem [{\citenamefont {Zacharias}\ \emph
  {et~al.}(2015{\natexlab{b}})\citenamefont {Zacharias}, \citenamefont
  {Rosch},\ and\ \citenamefont {Garst}}]{Zacharias2015-2}%
  \BibitemOpen
  \bibfield  {author} {\bibinfo {author} {\bibfnamefont {M.}~\bibnamefont
  {Zacharias}}, \bibinfo {author} {\bibfnamefont {A.}~\bibnamefont {Rosch}}, \
  and\ \bibinfo {author} {\bibfnamefont {M.}~\bibnamefont {Garst}},\ }\href
  {\doibase 10.1140/epjst/e2015-02444-5} {\bibfield  {journal} {\bibinfo
  {journal} {The European Physical Journal Special Topics}\ }\textbf {\bibinfo
  {volume} {224}},\ \bibinfo {pages} {1021} (\bibinfo {year}
  {2015}{\natexlab{b}})}\BibitemShut {NoStop}%
\bibitem [{\citenamefont {Villain}(1970)}]{Villain1970}%
  \BibitemOpen
  \bibfield  {author} {\bibinfo {author} {\bibfnamefont {J.}~\bibnamefont
  {Villain}},\ }\href {\doibase 10.1016/0038-1098(70)90453-9} {\bibfield
  {journal} {\bibinfo  {journal} {Solid State Commun.}\ }\textbf {\bibinfo
  {volume} {8}},\ \bibinfo {pages} {295} (\bibinfo {year} {1970})}\BibitemShut
  {NoStop}%
\bibitem [{\citenamefont {Levanyuk}\ and\ \citenamefont
  {Sobyanin}(1970)}]{Levanyuk1970}%
  \BibitemOpen
  \bibfield  {author} {\bibinfo {author} {\bibfnamefont {A.~P.}\ \bibnamefont
  {Levanyuk}}\ and\ \bibinfo {author} {\bibfnamefont {A.~A.}\ \bibnamefont
  {Sobyanin}},\ }\href@noop {} {\bibfield  {journal} {\bibinfo  {journal} {Sov.
  Phys. JETP Lett.}\ }\textbf {\bibinfo {volume} {11}},\ \bibinfo {pages} {371}
  (\bibinfo {year} {1970})}\BibitemShut {NoStop}%
\bibitem [{\citenamefont {Zacharias}\ \emph {et~al.}(2012)\citenamefont
  {Zacharias}, \citenamefont {Bartosch},\ and\ \citenamefont
  {Garst}}]{Zacharias2012}%
  \BibitemOpen
  \bibfield  {author} {\bibinfo {author} {\bibfnamefont {M.}~\bibnamefont
  {Zacharias}}, \bibinfo {author} {\bibfnamefont {L.}~\bibnamefont {Bartosch}},
  \ and\ \bibinfo {author} {\bibfnamefont {M.}~\bibnamefont {Garst}},\ }\href
  {\doibase 10.1103/PhysRevLett.109.176401} {\bibfield  {journal} {\bibinfo
  {journal} {Phys. Rev. Lett.}\ }\textbf {\bibinfo {volume} {109}},\ \bibinfo
  {pages} {176401} (\bibinfo {year} {2012})}\BibitemShut {NoStop}%
\bibitem [{\citenamefont {Gati}\ \emph {et~al.}(2016)\citenamefont {Gati},
  \citenamefont {Garst}, \citenamefont {Manna}, \citenamefont {Tutsch},
  \citenamefont {Wolf}, \citenamefont {Bartosch}, \citenamefont {Schubert},
  \citenamefont {Sasaki}, \citenamefont {Schlueter},\ and\ \citenamefont
  {Lang}}]{Gati2016}%
  \BibitemOpen
  \bibfield  {author} {\bibinfo {author} {\bibfnamefont {E.}~\bibnamefont
  {Gati}}, \bibinfo {author} {\bibfnamefont {M.}~\bibnamefont {Garst}},
  \bibinfo {author} {\bibfnamefont {R.~S.}\ \bibnamefont {Manna}}, \bibinfo
  {author} {\bibfnamefont {U.}~\bibnamefont {Tutsch}}, \bibinfo {author}
  {\bibfnamefont {B.}~\bibnamefont {Wolf}}, \bibinfo {author} {\bibfnamefont
  {L.}~\bibnamefont {Bartosch}}, \bibinfo {author} {\bibfnamefont
  {H.}~\bibnamefont {Schubert}}, \bibinfo {author} {\bibfnamefont
  {T.}~\bibnamefont {Sasaki}}, \bibinfo {author} {\bibfnamefont {J.~A.}\
  \bibnamefont {Schlueter}}, \ and\ \bibinfo {author} {\bibfnamefont
  {M.}~\bibnamefont {Lang}},\ }\href {\doibase 10.1126/sciadv.1601646}
  {\bibfield  {journal} {\bibinfo  {journal} {Science Advances}\ }\textbf
  {\bibinfo {volume} {2}},\ \bibinfo {pages} {e1601646} (\bibinfo {year}
  {2016})}\BibitemShut {NoStop}%
\bibitem [{\citenamefont {Weickert}\ \emph {et~al.}(2010)\citenamefont
  {Weickert}, \citenamefont {Brando}, \citenamefont {Steglich}, \citenamefont
  {Gegenwart},\ and\ \citenamefont {Garst}}]{Weickert2010}%
  \BibitemOpen
  \bibfield  {author} {\bibinfo {author} {\bibfnamefont {F.}~\bibnamefont
  {Weickert}}, \bibinfo {author} {\bibfnamefont {M.}~\bibnamefont {Brando}},
  \bibinfo {author} {\bibfnamefont {F.}~\bibnamefont {Steglich}}, \bibinfo
  {author} {\bibfnamefont {P.}~\bibnamefont {Gegenwart}}, \ and\ \bibinfo
  {author} {\bibfnamefont {M.}~\bibnamefont {Garst}},\ }\href {\doibase
  10.1103/PhysRevB.81.134438} {\bibfield  {journal} {\bibinfo  {journal} {Phys.
  Rev. B}\ }\textbf {\bibinfo {volume} {81}},\ \bibinfo {pages} {134438}
  (\bibinfo {year} {2010})}\BibitemShut {NoStop}%
\bibitem [{\citenamefont {Paul}\ and\ \citenamefont {Garst}(2017)}]{Paul2017}%
  \BibitemOpen
  \bibfield  {author} {\bibinfo {author} {\bibfnamefont {I.}~\bibnamefont
  {Paul}}\ and\ \bibinfo {author} {\bibfnamefont {M.}~\bibnamefont {Garst}},\
  }\href {\doibase 10.1103/PhysRevLett.118.227601} {\bibfield  {journal}
  {\bibinfo  {journal} {Phys. Rev. Lett.}\ }\textbf {\bibinfo {volume} {118}},\
  \bibinfo {pages} {227601} (\bibinfo {year} {2017})}\BibitemShut {NoStop}%
\bibitem [{\citenamefont {Reiss}\ \emph {et~al.}(2020)\citenamefont {Reiss},
  \citenamefont {Graf}, \citenamefont {Haghighirad}, \citenamefont {Knafo},
  \citenamefont {Drigo}, \citenamefont {Bristow}, \citenamefont {Schofield},\
  and\ \citenamefont {Coldea}}]{Reiss2020}%
  \BibitemOpen
  \bibfield  {author} {\bibinfo {author} {\bibfnamefont {P.}~\bibnamefont
  {Reiss}}, \bibinfo {author} {\bibfnamefont {D.}~\bibnamefont {Graf}},
  \bibinfo {author} {\bibfnamefont {A.~A.}\ \bibnamefont {Haghighirad}},
  \bibinfo {author} {\bibfnamefont {W.}~\bibnamefont {Knafo}}, \bibinfo
  {author} {\bibfnamefont {L.}~\bibnamefont {Drigo}}, \bibinfo {author}
  {\bibfnamefont {M.}~\bibnamefont {Bristow}}, \bibinfo {author} {\bibfnamefont
  {A.~J.}\ \bibnamefont {Schofield}}, \ and\ \bibinfo {author} {\bibfnamefont
  {A.~I.}\ \bibnamefont {Coldea}},\ }\href {\doibase 10.1038/s41567-019-0694-2}
  {\bibfield  {journal} {\bibinfo  {journal} {Nature Physics}\ }\textbf
  {\bibinfo {volume} {16}},\ \bibinfo {pages} {89} (\bibinfo {year}
  {2020})}\BibitemShut {NoStop}%
\bibitem [{\citenamefont {Rice}(1954)}]{Rice1954}%
  \BibitemOpen
  \bibfield  {author} {\bibinfo {author} {\bibfnamefont {O.~K.}\ \bibnamefont
  {Rice}},\ }\href {\doibase 10.1063/1.1740453} {\bibfield  {journal} {\bibinfo
   {journal} {J. Chem. Phys.}\ }\textbf {\bibinfo {volume} {22}},\ \bibinfo
  {pages} {1535} (\bibinfo {year} {1954})}\BibitemShut {NoStop}%
\bibitem [{\citenamefont {Domb}(1956)}]{Domb1956}%
  \BibitemOpen
  \bibfield  {author} {\bibinfo {author} {\bibfnamefont {C.}~\bibnamefont
  {Domb}},\ }\href {\doibase 10.1063/1.1743060} {\bibfield  {journal} {\bibinfo
   {journal} {J. Chem. Phys.}\ }\textbf {\bibinfo {volume} {25}},\ \bibinfo
  {pages} {783} (\bibinfo {year} {1956})}\BibitemShut {NoStop}%
\bibitem [{\citenamefont {Mattis}\ and\ \citenamefont
  {Schultz}(1963)}]{Mattis1963}%
  \BibitemOpen
  \bibfield  {author} {\bibinfo {author} {\bibfnamefont {D.~C.}\ \bibnamefont
  {Mattis}}\ and\ \bibinfo {author} {\bibfnamefont {T.~D.}\ \bibnamefont
  {Schultz}},\ }\href {\doibase 10.1103/PhysRev.129.175} {\bibfield  {journal}
  {\bibinfo  {journal} {Phys. Rev.}\ }\textbf {\bibinfo {volume} {129}},\
  \bibinfo {pages} {175} (\bibinfo {year} {1963})}\BibitemShut {NoStop}%
\bibitem [{\citenamefont {Fisher}(1968)}]{Fisher1968}%
  \BibitemOpen
  \bibfield  {author} {\bibinfo {author} {\bibfnamefont {M.~E.}\ \bibnamefont
  {Fisher}},\ }\href {\doibase 10.1103/PhysRev.176.257} {\bibfield  {journal}
  {\bibinfo  {journal} {Phys. Rev.}\ }\textbf {\bibinfo {volume} {176}},\
  \bibinfo {pages} {257} (\bibinfo {year} {1968})}\BibitemShut {NoStop}%
\bibitem [{\citenamefont {Larkin}\ and\ \citenamefont
  {Pikin}(1969)}]{LarkinPikin1969}%
  \BibitemOpen
  \bibfield  {author} {\bibinfo {author} {\bibfnamefont {A.}~\bibnamefont
  {Larkin}}\ and\ \bibinfo {author} {\bibfnamefont {S.}~\bibnamefont {Pikin}},\
  }\href {http://www.tcm.phy.cam.ac.uk/~dek12/courses/phase/Ap2.pdf} {\bibfield
   {journal} {\bibinfo  {journal} {Sov. Phys. JETP}\ }\textbf {\bibinfo
  {volume} {29}},\ \bibinfo {pages} {891} (\bibinfo {year} {1969})}\BibitemShut
  {NoStop}%
\bibitem [{\citenamefont {Rudnick}\ \emph {et~al.}(1974)\citenamefont
  {Rudnick}, \citenamefont {Bergman},\ and\ \citenamefont
  {Imry}}]{Rudnick1974}%
  \BibitemOpen
  \bibfield  {author} {\bibinfo {author} {\bibfnamefont {J.}~\bibnamefont
  {Rudnick}}, \bibinfo {author} {\bibfnamefont {D.}~\bibnamefont {Bergman}}, \
  and\ \bibinfo {author} {\bibfnamefont {Y.}~\bibnamefont {Imry}},\ }\href
  {\doibase 10.1016/0375-9601(74)90959-1} {\bibfield  {journal} {\bibinfo
  {journal} {Phys. Lett. A}\ }\textbf {\bibinfo {volume} {46}},\ \bibinfo
  {pages} {449} (\bibinfo {year} {1974})}\BibitemShut {NoStop}%
\bibitem [{\citenamefont {Sak}(1974)}]{Sak1974}%
  \BibitemOpen
  \bibfield  {author} {\bibinfo {author} {\bibfnamefont {J.}~\bibnamefont
  {Sak}},\ }\href {\doibase 10.1103/PhysRevB.10.3957} {\bibfield  {journal}
  {\bibinfo  {journal} {Phys. Rev. B}\ }\textbf {\bibinfo {volume} {10}},\
  \bibinfo {pages} {3957} (\bibinfo {year} {1974})}\BibitemShut {NoStop}%
\bibitem [{\citenamefont {Wegner}(1974)}]{Wegner1974}%
  \BibitemOpen
  \bibfield  {author} {\bibinfo {author} {\bibfnamefont {F.}~\bibnamefont
  {Wegner}},\ }\href {\doibase 10.1088/0022-3719/7/12/005} {\bibfield
  {journal} {\bibinfo  {journal} {J. Phys. C}\ }\textbf {\bibinfo {volume}
  {7}},\ \bibinfo {pages} {2109} (\bibinfo {year} {1974})}\BibitemShut
  {NoStop}%
\bibitem [{\citenamefont {Bergman}\ and\ \citenamefont
  {Halperin}(1976)}]{BergmanHalperin1976}%
  \BibitemOpen
  \bibfield  {author} {\bibinfo {author} {\bibfnamefont {D.~J.}\ \bibnamefont
  {Bergman}}\ and\ \bibinfo {author} {\bibfnamefont {B.~I.}\ \bibnamefont
  {Halperin}},\ }\href {\doibase 10.1103/PhysRevB.13.2145} {\bibfield
  {journal} {\bibinfo  {journal} {Phys. Rev. B}\ }\textbf {\bibinfo {volume}
  {13}},\ \bibinfo {pages} {2145} (\bibinfo {year} {1976})}\BibitemShut
  {NoStop}%
\bibitem [{\citenamefont {De~Moura}\ \emph {et~al.}(1976)\citenamefont
  {De~Moura}, \citenamefont {Lubensky}, \citenamefont {Imry},\ and\
  \citenamefont {Aharony}}]{DeMoura1976}%
  \BibitemOpen
  \bibfield  {author} {\bibinfo {author} {\bibfnamefont {M.~A.}\ \bibnamefont
  {De~Moura}}, \bibinfo {author} {\bibfnamefont {T.~C.}\ \bibnamefont
  {Lubensky}}, \bibinfo {author} {\bibfnamefont {Y.}~\bibnamefont {Imry}}, \
  and\ \bibinfo {author} {\bibfnamefont {A.}~\bibnamefont {Aharony}},\ }\href
  {\doibase 10.1103/PhysRevB.13.2176} {\bibfield  {journal} {\bibinfo
  {journal} {Phys. Rev. B}\ }\textbf {\bibinfo {volume} {13}},\ \bibinfo
  {pages} {2176} (\bibinfo {year} {1976})}\BibitemShut {NoStop}%
\bibitem [{\citenamefont {Nattermann}(1977)}]{Nattermann1977}%
  \BibitemOpen
  \bibfield  {author} {\bibinfo {author} {\bibfnamefont {T.}~\bibnamefont
  {Nattermann}},\ }\href@noop {} {\bibfield  {journal} {\bibinfo  {journal} {J.
  Phys. A}\ }\textbf {\bibinfo {volume} {10}},\ \bibinfo {pages} {757}
  (\bibinfo {year} {1977})}\BibitemShut {NoStop}%
\bibitem [{\citenamefont {Bruno}\ and\ \citenamefont
  {Sak}(1980)}]{BrunoSak1980}%
  \BibitemOpen
  \bibfield  {author} {\bibinfo {author} {\bibfnamefont {J.}~\bibnamefont
  {Bruno}}\ and\ \bibinfo {author} {\bibfnamefont {J.}~\bibnamefont {Sak}},\
  }\href {\doibase 10.1103/PhysRevB.22.3302} {\bibfield  {journal} {\bibinfo
  {journal} {Phys. Rev. B}\ }\textbf {\bibinfo {volume} {22}},\ \bibinfo
  {pages} {3302} (\bibinfo {year} {1980})}\BibitemShut {NoStop}%
\bibitem [{\citenamefont {D\"unweg}(2000)}]{Duenweg2000}%
  \BibitemOpen
  \bibfield  {author} {\bibinfo {author} {\bibfnamefont {B.}~\bibnamefont
  {D\"unweg}},\ }\href
  {https://www2.mpip-mainz.mpg.de/~duenweg/Public/PDFsOfPreprints/habil.pdf}
  {\bibfield  {journal} {\bibinfo  {journal} {Habilitationsschrift}\ }
  (\bibinfo {year} {2000})}\BibitemShut {NoStop}%
\bibitem [{\citenamefont {Landau}\ \emph {et~al.}(1986)\citenamefont {Landau},
  \citenamefont {Pitaevskii}, \citenamefont {Lifshitz},\ and\ \citenamefont
  {Kosevich}}]{Landau1986}%
  \BibitemOpen
  \bibfield  {author} {\bibinfo {author} {\bibfnamefont {L.~D.}\ \bibnamefont
  {Landau}}, \bibinfo {author} {\bibfnamefont {L.~P.}\ \bibnamefont
  {Pitaevskii}}, \bibinfo {author} {\bibfnamefont {E.~M.}\ \bibnamefont
  {Lifshitz}}, \ and\ \bibinfo {author} {\bibfnamefont {A.~M.}\ \bibnamefont
  {Kosevich}},\ }\href@noop {} {\emph {\bibinfo {title} {Theory of
  Elasticity}}},\ \bibinfo {edition} {3rd}\ ed.\ (\bibinfo  {publisher}
  {Butterworth-Heinemann},\ \bibinfo {year} {1986})\BibitemShut {NoStop}%
\bibitem [{Note1()}]{Note1}%
  \BibitemOpen
  \bibinfo {note} {In Refs.~\protect \rev@citealpnum {Sak1974,BrunoSak1980}
  only free boundary conditions at constant $P$ were considered and the FR-WF
  fixed point was inaccessible.}\BibitemShut {Stop}%
\bibitem [{\citenamefont {Anfuso}\ \emph {et~al.}(2008)\citenamefont {Anfuso},
  \citenamefont {Garst}, \citenamefont {Rosch}, \citenamefont {Heyer},
  \citenamefont {Lorenz}, \citenamefont {R\"uegg},\ and\ \citenamefont
  {Kr\"amer}}]{Anfuso2008}%
  \BibitemOpen
  \bibfield  {author} {\bibinfo {author} {\bibfnamefont {F.}~\bibnamefont
  {Anfuso}}, \bibinfo {author} {\bibfnamefont {M.}~\bibnamefont {Garst}},
  \bibinfo {author} {\bibfnamefont {A.}~\bibnamefont {Rosch}}, \bibinfo
  {author} {\bibfnamefont {O.}~\bibnamefont {Heyer}}, \bibinfo {author}
  {\bibfnamefont {T.}~\bibnamefont {Lorenz}}, \bibinfo {author} {\bibfnamefont
  {C.}~\bibnamefont {R\"uegg}}, \ and\ \bibinfo {author} {\bibfnamefont
  {K.}~\bibnamefont {Kr\"amer}},\ }\href {\doibase 10.1103/PhysRevB.77.235113}
  {\bibfield  {journal} {\bibinfo  {journal} {Phys. Rev. B}\ }\textbf {\bibinfo
  {volume} {77}},\ \bibinfo {pages} {235113} (\bibinfo {year}
  {2008})}\BibitemShut {NoStop}%
\bibitem [{\citenamefont {Noad}\ \emph {et~al.}(2023)\citenamefont {Noad},
  \citenamefont {Ishida}, \citenamefont {Li}, \citenamefont {Gati},
  \citenamefont {Stangier}, \citenamefont {Kikugawa}, \citenamefont {Sokolov},
  \citenamefont {Nicklas}, \citenamefont {Kim}, \citenamefont {Mazin},
  \citenamefont {Garst}, \citenamefont {Schmalian}, \citenamefont {Mackenzie},\
  and\ \citenamefont {Hicks}}]{Noad2023}%
  \BibitemOpen
  \bibfield  {author} {\bibinfo {author} {\bibfnamefont {H.~M.~L.}\
  \bibnamefont {Noad}}, \bibinfo {author} {\bibfnamefont {K.}~\bibnamefont
  {Ishida}}, \bibinfo {author} {\bibfnamefont {Y.-S.}\ \bibnamefont {Li}},
  \bibinfo {author} {\bibfnamefont {E.}~\bibnamefont {Gati}}, \bibinfo {author}
  {\bibfnamefont {V.~C.}\ \bibnamefont {Stangier}}, \bibinfo {author}
  {\bibfnamefont {N.}~\bibnamefont {Kikugawa}}, \bibinfo {author}
  {\bibfnamefont {D.~A.}\ \bibnamefont {Sokolov}}, \bibinfo {author}
  {\bibfnamefont {M.}~\bibnamefont {Nicklas}}, \bibinfo {author} {\bibfnamefont
  {B.}~\bibnamefont {Kim}}, \bibinfo {author} {\bibfnamefont {I.~I.}\
  \bibnamefont {Mazin}}, \bibinfo {author} {\bibfnamefont {M.}~\bibnamefont
  {Garst}}, \bibinfo {author} {\bibfnamefont {J.}~\bibnamefont {Schmalian}},
  \bibinfo {author} {\bibfnamefont {A.~P.}\ \bibnamefont {Mackenzie}}, \ and\
  \bibinfo {author} {\bibfnamefont {C.~W.}\ \bibnamefont {Hicks}},\ }\href@noop
  {} {\enquote {\bibinfo {title} {Giant lattice softening at a {Lifshitz}
  transition in {Sr$_{2}$RuO$_{4}$}},}\ } (\bibinfo {year} {2023}),\ \Eprint
  {http://arxiv.org/abs/2306.17835} {arXiv:2306.17835} \BibitemShut {NoStop}%
\bibitem [{\citenamefont {Chandra}\ \emph {et~al.}(2020)\citenamefont
  {Chandra}, \citenamefont {Coleman}, \citenamefont {Continentino},\ and\
  \citenamefont {Lonzarich}}]{Chandra2020}%
  \BibitemOpen
  \bibfield  {author} {\bibinfo {author} {\bibfnamefont {P.}~\bibnamefont
  {Chandra}}, \bibinfo {author} {\bibfnamefont {P.}~\bibnamefont {Coleman}},
  \bibinfo {author} {\bibfnamefont {M.~A.}\ \bibnamefont {Continentino}}, \
  and\ \bibinfo {author} {\bibfnamefont {G.~G.}\ \bibnamefont {Lonzarich}},\
  }\href {\doibase 10.1103/PhysRevResearch.2.043440} {\bibfield  {journal}
  {\bibinfo  {journal} {Phys. Rev. Res.}\ }\textbf {\bibinfo {volume} {2}},\
  \bibinfo {pages} {043440} (\bibinfo {year} {2020})}\BibitemShut {NoStop}%
\bibitem [{\citenamefont {Samanta}\ \emph {et~al.}(2022)\citenamefont
  {Samanta}, \citenamefont {Shimshoni},\ and\ \citenamefont
  {Podolsky}}]{Samanta2022}%
  \BibitemOpen
  \bibfield  {author} {\bibinfo {author} {\bibfnamefont {A.}~\bibnamefont
  {Samanta}}, \bibinfo {author} {\bibfnamefont {E.}~\bibnamefont {Shimshoni}},
  \ and\ \bibinfo {author} {\bibfnamefont {D.}~\bibnamefont {Podolsky}},\
  }\href {\doibase 10.1103/PhysRevB.106.035154} {\bibfield  {journal} {\bibinfo
   {journal} {Phys. Rev. B}\ }\textbf {\bibinfo {volume} {106}},\ \bibinfo
  {pages} {035154} (\bibinfo {year} {2022})}\BibitemShut {NoStop}%
\bibitem [{\citenamefont {Zhu}\ \emph {et~al.}(2003)\citenamefont {Zhu},
  \citenamefont {Garst}, \citenamefont {Rosch},\ and\ \citenamefont
  {Si}}]{Zhu2003}%
  \BibitemOpen
  \bibfield  {author} {\bibinfo {author} {\bibfnamefont {L.}~\bibnamefont
  {Zhu}}, \bibinfo {author} {\bibfnamefont {M.}~\bibnamefont {Garst}}, \bibinfo
  {author} {\bibfnamefont {A.}~\bibnamefont {Rosch}}, \ and\ \bibinfo {author}
  {\bibfnamefont {Q.}~\bibnamefont {Si}},\ }\href {\doibase
  10.1103/PhysRevLett.91.066404} {\bibfield  {journal} {\bibinfo  {journal}
  {Phys. Rev. Lett.}\ }\textbf {\bibinfo {volume} {91}},\ \bibinfo {pages}
  {066404} (\bibinfo {year} {2003})}\BibitemShut {NoStop}%
\bibitem [{\citenamefont {Garst}\ and\ \citenamefont
  {Rosch}(2005)}]{Garst2005}%
  \BibitemOpen
  \bibfield  {author} {\bibinfo {author} {\bibfnamefont {M.}~\bibnamefont
  {Garst}}\ and\ \bibinfo {author} {\bibfnamefont {A.}~\bibnamefont {Rosch}},\
  }\href {\doibase 10.1103/PhysRevB.72.205129} {\bibfield  {journal} {\bibinfo
  {journal} {Phys. Rev. B}\ }\textbf {\bibinfo {volume} {72}},\ \bibinfo
  {pages} {205129} (\bibinfo {year} {2005})}\BibitemShut {NoStop}%
\bibitem [{\citenamefont {Sitte}\ \emph {et~al.}(2009)\citenamefont {Sitte},
  \citenamefont {Rosch}, \citenamefont {Meyer}, \citenamefont {Matveev},\ and\
  \citenamefont {Garst}}]{Sitte2009}%
  \BibitemOpen
  \bibfield  {author} {\bibinfo {author} {\bibfnamefont {M.}~\bibnamefont
  {Sitte}}, \bibinfo {author} {\bibfnamefont {A.}~\bibnamefont {Rosch}},
  \bibinfo {author} {\bibfnamefont {J.~S.}\ \bibnamefont {Meyer}}, \bibinfo
  {author} {\bibfnamefont {K.~A.}\ \bibnamefont {Matveev}}, \ and\ \bibinfo
  {author} {\bibfnamefont {M.}~\bibnamefont {Garst}},\ }\href {\doibase
  10.1103/PhysRevLett.102.176404} {\bibfield  {journal} {\bibinfo  {journal}
  {Phys. Rev. Lett.}\ }\textbf {\bibinfo {volume} {102}},\ \bibinfo {pages}
  {176404} (\bibinfo {year} {2009})}\BibitemShut {NoStop}%
\bibitem [{\citenamefont {Merchant}\ \emph {et~al.}(2014)\citenamefont
  {Merchant}, \citenamefont {Normand}, \citenamefont {Kr{\"a}mer},
  \citenamefont {Boehm}, \citenamefont {McMorrow},\ and\ \citenamefont
  {R{\"u}egg}}]{Merchant2014}%
  \BibitemOpen
  \bibfield  {author} {\bibinfo {author} {\bibfnamefont {P.}~\bibnamefont
  {Merchant}}, \bibinfo {author} {\bibfnamefont {B.}~\bibnamefont {Normand}},
  \bibinfo {author} {\bibfnamefont {K.~W.}\ \bibnamefont {Kr{\"a}mer}},
  \bibinfo {author} {\bibfnamefont {M.}~\bibnamefont {Boehm}}, \bibinfo
  {author} {\bibfnamefont {D.~F.}\ \bibnamefont {McMorrow}}, \ and\ \bibinfo
  {author} {\bibfnamefont {C.}~\bibnamefont {R{\"u}egg}},\ }\href {\doibase
  10.1038/nphys2902} {\bibfield  {journal} {\bibinfo  {journal} {Nature
  Physics}\ }\textbf {\bibinfo {volume} {10}},\ \bibinfo {pages} {373}
  (\bibinfo {year} {2014})}\BibitemShut {NoStop}%
\end{thebibliography}%
\end{document}